\documentclass{aa}  

\usepackage{graphicx}
\usepackage{pdflscape}
\usepackage{txfonts}
\usepackage{placeins}
\usepackage{hyperref}
\hypersetup{colorlinks=true,citecolor=blue,linkcolor=black, urlcolor=black, linktoc=all}
\usepackage{subcaption}

\newcommand{\cyc}[1]{c-C$_3$H$_2$} 
\newcommand{\meth}[1]{CH$_3$OH}
\newcommand{\prop}[1]{CH$_3$CCH}
\newcommand{\dprop}[1]{CH$_2$DCCH}
\newcommand{\propd}[1]{CH$_3$CCD}
\newcommand{\cyano}[1]{HC$_3$N}
\newcommand{\xcyano}[1]{HCC$^{13}$CN}
\newcommand{\dcyano}[1]{DC$_3$N}

\begin{document}

   \title{Deuteration of \cyano{} and \prop{} in the pre-stellar core L1544}

   \author{K. Giers
          \inst{1},
          S. Spezzano\inst{1}, 
          Y. Lin\inst{1},
          P. Caselli\inst{1},
          O. Sipil\"{a}\inst{1}
          }

   \institute{Max-Planck-Institute for Extraterrestrial Physics, Giessenbachstrasse 1, D-85748 Garching, Germany\\
              \email{kgiers@mpe.mpg.de}
         }


 
  \abstract
   {Deuterated molecules are a useful diagnostic tool to probe the evolution and the kinematics in the earliest stages of star formation. Due to the low temperatures and high densities in the centre of pre-stellar cores, the deuterium fraction is enhanced by several orders of magnitude with respect to the cosmic D/H abundance ratio. }
   {We study the distribution of the emission and the deuteration of the two carbon chains \cyano{} and \prop{} throughout the prototypical pre-stellar core L1544. } 
   {We analyse emission maps of \prop{}, \dprop{}, \propd{}, \cyano{}, \xcyano{}, and \dcyano{}, observed towards L1544 with the IRAM 30\,m single-dish radio telescope. 
   We use non-local thermodynamic equilibrium radiative transfer calculations, combined with chemical modelling of the molecular abundances, to constrain physical parameters of the observed species. 
   Following this, we derive the corresponding column density and deuteration maps and analyse the chemical processes influencing the different molecular distributions.}
   {We find levels of deuteration of N(\dcyano{})/N(\cyano{}) = $0.04-0.07$, N(\dprop{})/N(\prop{}) = $0.09-0.15$, and N(\propd{})/N(\prop{}) = $0.07-0.09$. 
   The deuteration of \cyano{} appears homogeneous across the core, with widespread D-fraction values above 0.06, 
   tracing intermediate-density gas in the outer layers of the core, at densities less than $\rm 10^5\,cm^{-3}$. 
   \propd{} is most efficiently formed in the higher-density regions towards the core centre, while the deuteration fraction of \dprop{} traces a local density enhancement in the north-east of the core, coinciding with the \meth{} emission peak.
   }
   {The results suggest that gas-phase reactions dominate the formation and deuteration of both \cyano{} and \prop{} in L1544, with spatial variations driven by physical structure, density and external radiation. 
   The significantly higher deuteration fraction of \dprop{} compared to \propd{} and a tentative gradient with higher values in the north suggest that there are different deuteration mechanisms for the two functional groups, with varying efficiency across the core.
   Similarities between the \dprop{} emission and CH$_2$DOH might indicate an additional deuteration pathway of \prop{} on the surfaces of dust grains, as observed for H$_2$CO.
   }

               
\authorrunning{Giers et al.}

   \maketitle
%


\section{Introduction}

Deuterium fractionation of molecules is a powerful tool to study the early stages of star formation.
Due to the combination of high densities and low temperatures in the pre-stellar phase, CO is largely depleted onto the surfaces of dust grains \citep[see e.g.][]{Caselli1999}. 
As CO is the main destroyer of several ions (e.g. H$_3^+$ and its deuterated forms; \citealt{DalgarnoLepp1984}), its depletion leads to an enhancement in their abundances. Simultaneously, the main pathway for the formation of H$_2$D$^+$ opens up:
\begin{equation}
    \text{H}_3^+ + \text{HD} \longrightarrow \text{H}_2\text{D}^+ + \text{H}_2 + 232\,\text{K}\;.
\end{equation}
As this reaction is exothermic, the backwards reaction is suppressed at the low temperatures present in pre-stellar cores (assuming a low ortho-to-para H$_2$ ratio; e.g. \citealt{Pagani1992}).
The enhanced abundance of H$_2$D$^+$, as well as of $\rm D_2H^+$ and D$_3^+$, then leads to an increased deuteration of other molecules, via a deuteron transfer, as predicted by models \citep[e.g.][]{Walmsley2004}.

The evolved pre-stellar core L1544 is known for its high levels of deuteration \citep[see e.g.][]{Crapsi2005,Chantzos2018,Giers2023}. The distribution of the deuterium fractionation across this core has been studied for a variety of molecules: HCO$^+$, N$_2$H$^+$, \meth{}, H$_2$CO, \cyc{}, and H$_2$CS \citep{Caselli2002a,Redaelli2019,ChaconTanarro2019,Giers2022,Spezzano2022}.
Those observations show that, for most species, the deuterium fractionation in L1544 is most efficient in the central regions of the core, indicated by the dust continuum peak, in higher-density gas layers.
On the other hand, the different molecules show a wide range of deuteration levels: while H$_2$CO and HCO$^+$ show rather low values of around 3\%, N$_2$H$^+$ and H$_2$CS reach high values of up to 30\%. 
In addition to its chemical richness, the core shows a spatial molecular segregation with characteristic emission peaks of molecules, namely \cyc{}, \meth{}, and HNCO \citep{Spezzano2017}. 
These seem to be the emission peaks of several other species that are chemically related to the respective molecules: carbon-chain molecules are grouped in the south on the \cyc{} peak, O-bearing molecules such as \meth{} or SO gather in the north-east on the \meth{} peak, and HNCO in the north-west is joined by \prop{}. 
The spatial differentiation between carbon chains in the south and \meth{} in the north has been explained by a non-uniform external illumination of the core \citep{Spezzano2016}. 
More precisely, L1544 is located at the edge of a filament in the Taurus molecular cloud, which leads to the southern part being exposed to the interstellar radiation field (ISRF), while the northern part is more protected by the molecular cloud. The ISRF drives the photochemistry in the south, enhancing the abundance of free carbon atoms in the gas phase, and subsequently the abundance of carbon chains. In the protected north, however, photochemistry is not a dominant process, and carbon is mainly locked in CO. 
The northern part of the core is also known to be the meeting point of two filamentary structures \citep[e.g. see][]{Spezzano2016,Lin2022}. \cite{Giers2025} suggest that inflowing, fresh material actually enhances the abundances of chemically young species in this area, causing the emission peak of \prop{} in the north-west of the core.

This work focuses on the emission and the distribution of the two carbon chains \cyano{} and \prop{} and their singly deuterated isotopologues. 
In L1544, the two molecules show a spatial differentiation (see Fig.~\ref{fig:IntegratedIntensityMaps}), where the emission of \cyano{} follows other carbon chains and peaks in the south of the core, while \prop{} peaks in the north-west on the HNCO peak.
Therefore, the observations of \cyano{}, \prop{}, and isotopologues offer complementary insights into the complex chemical structure of the prototypical pre-stellar core L1544. 
In addition, this dataset gives the opportunity to study the deuteration of larger molecules in L1544 in more detail, providing a deeper understanding of the chemical processes that dominate in pre-stellar cores and shape the observed chemical structures. 
In addition, the study of deuteration can help to better understand the formation pathways of the molecules and, eventually, put constrains on chemical models.

Cyanoacetylene, \cyano{}, is the smallest molecule from the family of cyanopolyynes, which are possible precursors of prebiotic molecules \citep[see e.g.][]{Shingledecker2021}. After its first detection in space by \cite{Turner1971}, \cyano{} has since been detected in various interstellar environments \citep[see e.g.][]{Morris1976,Walmsley1980,Chapillon2012,Spezzano2017}.
The deuterated counterpart, \dcyano{}, was first detected by \cite{Langer1980} towards the low-mass dark cloud TMC-1, and by \cite{Howe1994} towards several other dark cores.
Methylacetylene, \prop{}, is a large carbon chain, and a proposed precursor in the formation of polycyclic aromatic hydrocarbons (PAHs) \citep{ParkerKaiser2017}. 
It contains a methyl group with three equivalent H atoms and one H atom in the terminal CCH group, which makes the study of its deuteration very useful for understanding the molecule's formation pathways.
In the interstellar medium (ISM), \prop{} was first detected by \cite{BuhlSnyder1973}, followed by detections in various other environments \citep{vanDishoeck1995,Fontani2002,Vastel2014, Gratier2016, Spezzano2017, Agundez2019,LinWyrowski2022}.
The singly deuterated isotopologues \dprop{} and \propd{} were first detected towards the cold dark cloud TMC-1 \citep{Gerin1992,Markwick2005}. The detection of the doubly deuterated forms CHD$_2$CCH and CH$_2$DCCD was reported by \cite{Agundez2021}.

Both \cyano{} and \prop{} are believed to primarily form in the gas phase. 
The main formation pathways for \prop{} are thought to involve ion-molecule reactions, neutral-neutral reactions, and dissociative recombination of larger hydrocarbons \citep{SchiffBohme1979,Turner1999,Calcutt2019,Giers2025}, while for \cyano{}, potential formation routes are via HCN, CN, or HNC (\citealt{HCNroute,CNroute,HNCroute}; see \citealt{HilyBlant2018b} for an overview).
Not much is known about the deuteration process of either species, but there is evidence that it occurs mainly in the gas phase.
According to \cite{Rivilla2020}, \dcyano{} most likely forms via ion-molecule reactions, for example of \cyano{} directly reacting with H$_2$D$^+$, whereas the deuteration of \prop{} is believed to occur via the dissociative recombination of $\rm C_3H_6D^+$ and $\rm C_3H_5D^+$ \citep{Agundez2021}.

In this work, we present the emission maps of \dprop{}, \propd{}, and \dcyano{}, and their non-deuterated isotopologues \prop{}, \cyano{}, and \xcyano{}, observed towards the pre-stellar core L1544. 
We report the observations and detected lines in Sect.~\ref{sec:data}.
In Sect.~\ref{sec:LOCmodelling}, we model spectral lines extracted towards the dust peak of the source with non-local thermodynamic equilibrium (non-LTE) radiative transfer calculations to constrain physical parameters of the molecular emission.
Following this, we derive the corresponding column density and deuterium fraction maps of the different species in Sect.~\ref{sec:ColDensandDeutMaps}, assuming LTE conditions.
In Sect.~\ref{sec:discussion}, we discuss the deuterium fractionation of \cyano{} and \prop{} in L1544 and the implications for the understanding of the ongoing chemical processes.
Section~\ref{sec:conclusion} concludes the article.

\section{Data}\label{sec:data}

The data presented in this work were taken with the IRAM 30\,m single-dish radio telescope on Pico Veleta in the Sierra Nevada, Spain. 
The observations were carried out between October 2013 and June 2022 (PIs: Silvia Spezzano, Katharina Giers).
The data for \cyano{}, \xcyano{}, \prop{}, and \dprop{} have also been used in \cite{Spezzano2017} and \cite{Giers2025}.
The on-the-fly (OTF) maps were observed in position switching mode, using the EMIR E090 receiver and the Fourier transform spectrometer (FTS) backend with a spectral resolution of 50\,kHz.
The observed $2.5'\times2.5'$ OTF maps were centred on the source dust emission peak ($\alpha_{2000}=05^\mathrm{h}04^\mathrm{m}17^\mathrm{s}.21$, $\delta_{2000}=+25^\circ10'42''.8$, \citealt{WardThompson1999}).
The observed transitions are summarised in Table~\ref{Tab:ObservedLinesAll}.

\begin{figure*}[h]
    \centering
    \includegraphics[width=0.33\textwidth]{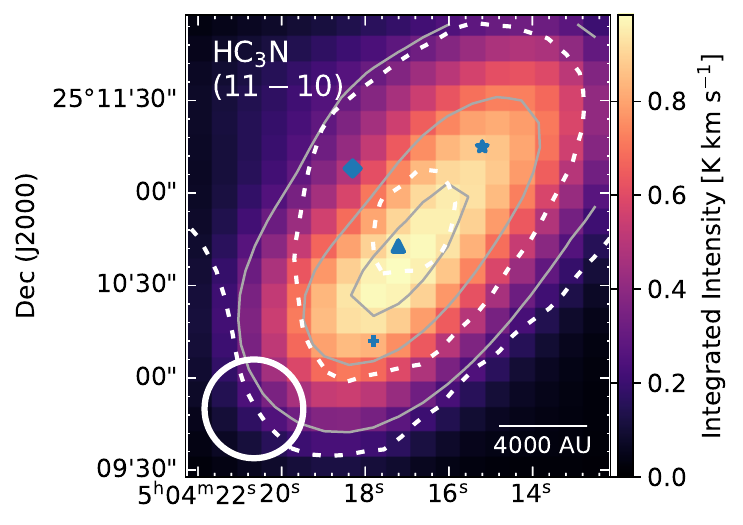}
    \includegraphics[width=0.33\textwidth]{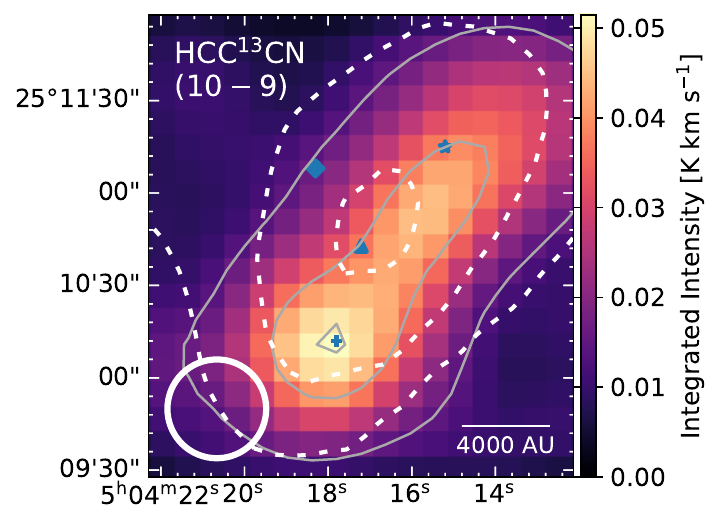}
    \includegraphics[width=0.33\textwidth]{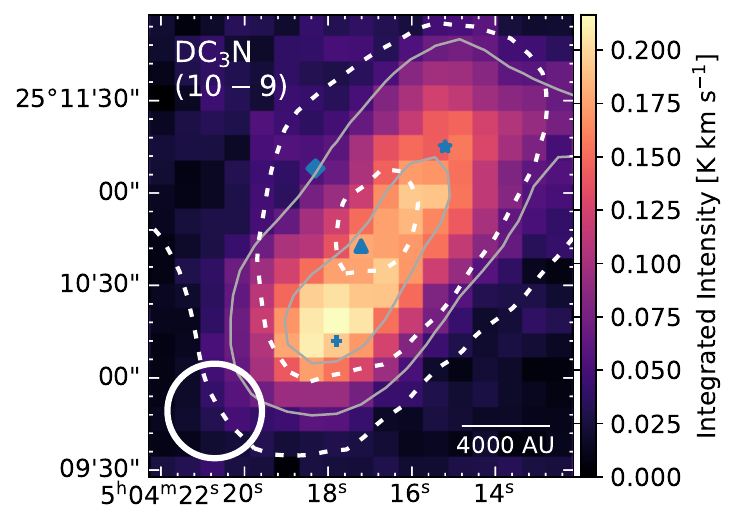}
    \includegraphics[width=0.33\textwidth]{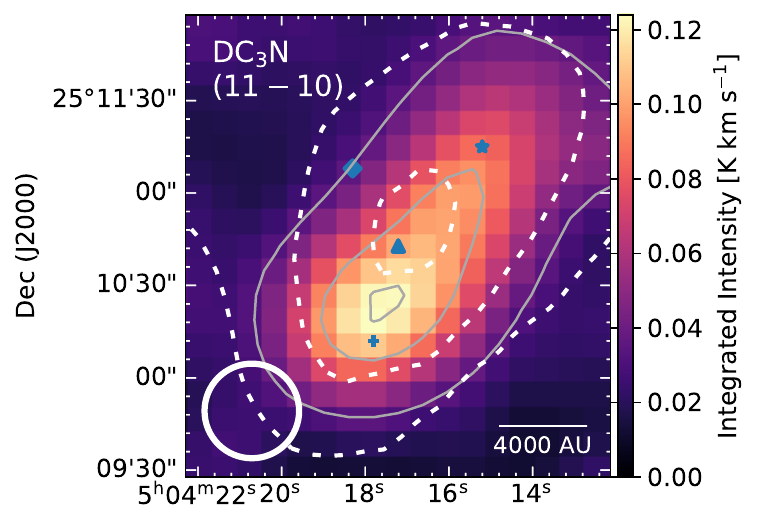}
    \includegraphics[width=0.33\textwidth]{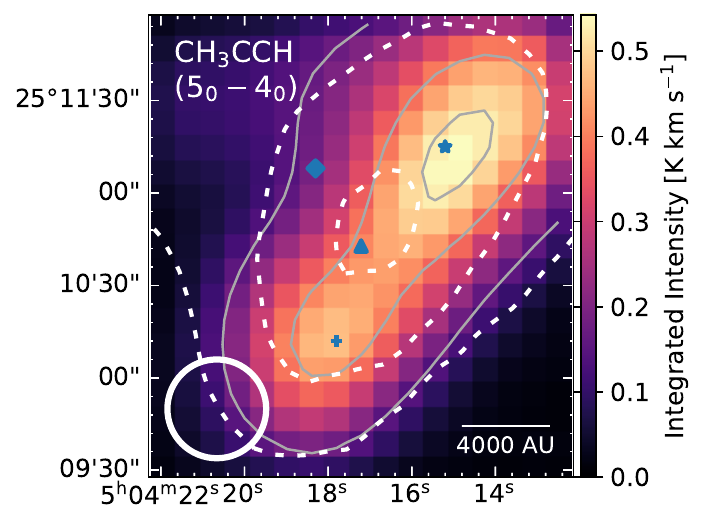}
    \includegraphics[width=0.33\textwidth]{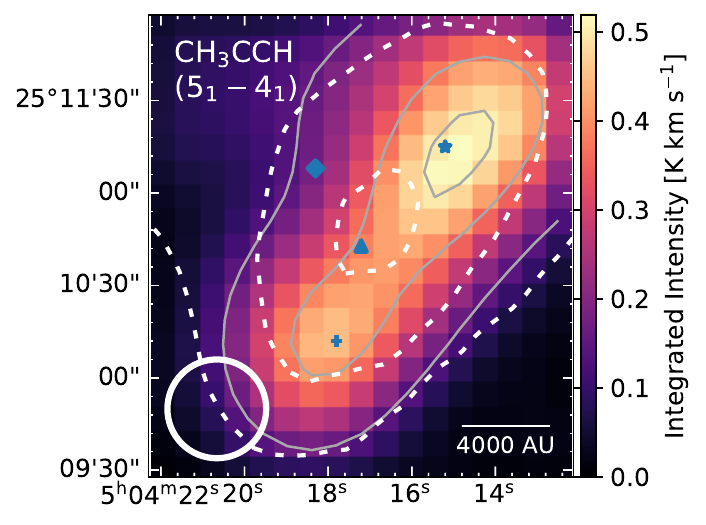}
    \includegraphics[width=0.33\textwidth]{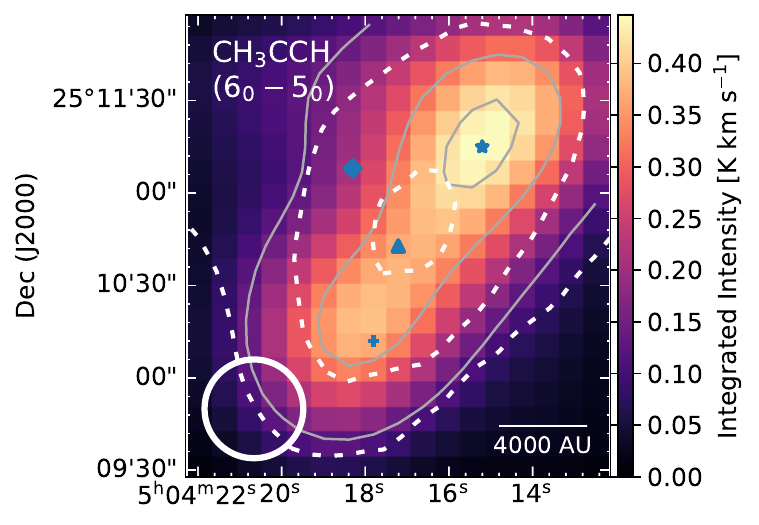}
    \includegraphics[width=0.33\textwidth]{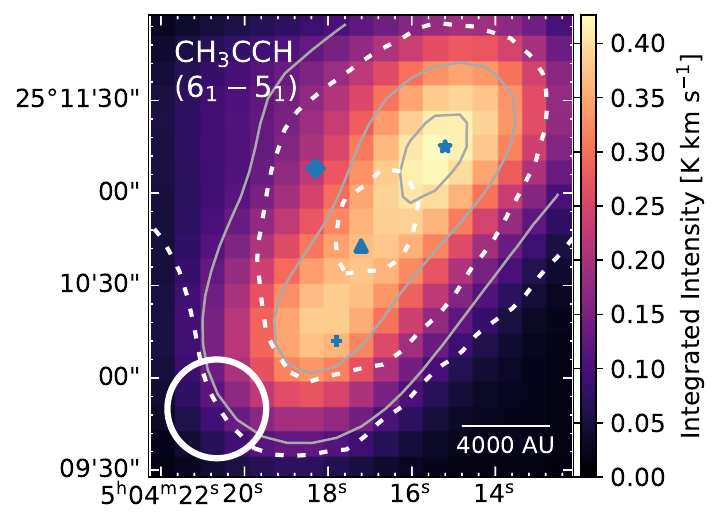}
    \includegraphics[width=0.33\textwidth]{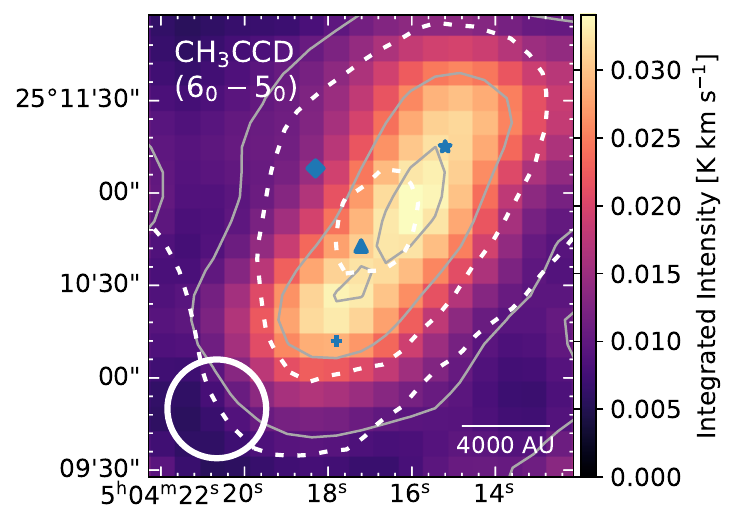}
    \includegraphics[width=0.33\textwidth]{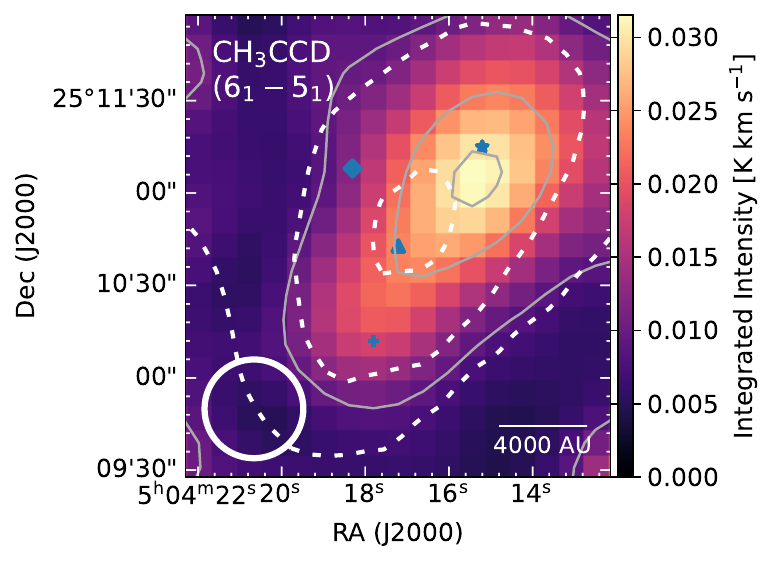}
    \includegraphics[width=0.33\textwidth]{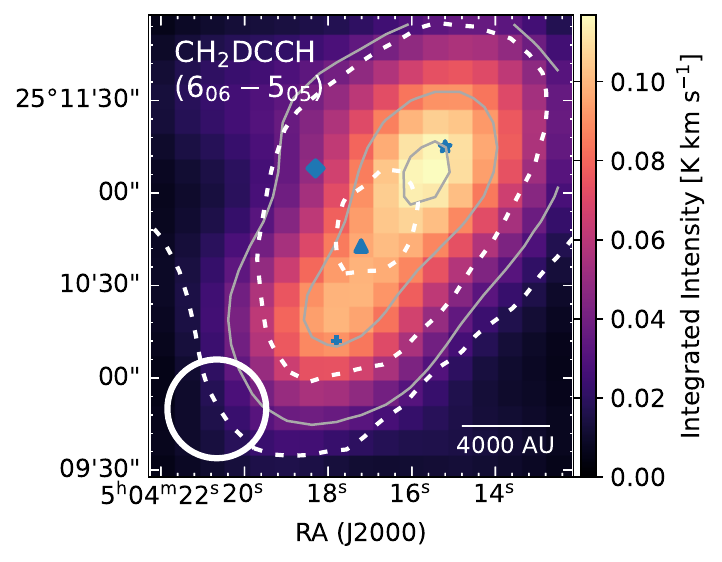}
    \caption{Integrated intensity maps of the observed transitions. The grey solid line contours indicate the 30\%, 70\%, and 90\% level of the peak integrated intensity. The dashed line contours represent 30\%, 50\%, and 90\% of the H$_2$ column density peak derived from $Herschel$ maps \citep{Spezzano2016}. The markers in blue represent the dust peak (triangle) and the molecular emission peaks of \meth{} (diamond), \prop{} (star), and \cyc{} (plus sign), where emission spectra (shown in Fig.~\ref{fig:CH3CCHmoleculepeaks}) are extracted within a circular aperture with a diameter corresponding to the telescope beam.
    The white circle in the bottom-left corner indicates the beam size of the IRAM 30\,m telescope (31").}
    \label{fig:IntegratedIntensityMaps}
\end{figure*}

Data processing was done using the GILDAS software \citep{Pety2005} and the python packages \texttt{pandas} \citep{pandas} and \texttt{spectral-cube} \citep{Spectralcube}.
All emission maps were gridded to a pixel size of 8" with the CLASS software in the GILDAS packages.
This corresponds to one-third to one-quarter of the actual beam size, depending on the frequency. To create a uniform dataset, we additionally resampled the data
to a spectral resolution of 0.18\,km\,s$^{-1}$, corresponding to the resolution of the lowest frequency observation (84.4\,GHz).
The antenna temperature $T_A^*$ was converted to the main beam temperature $T_\mathrm{mb}$ using the relation $T_\mathrm{mb}=F_\mathrm{eff}/B_\mathrm{eff}\cdot T_A^*$. The corresponding values for the 30\,m forward ($F_\mathrm{eff}$) and main-beam efficiencies ($B_\mathrm{eff}$) are given in Table~\ref{Tab:ObservedLinesAll}.
Figure~\ref{fig:IntegratedIntensityMaps} shows the integrated intensity maps of the observed transitions, computed by integrating over a velocity range of 0.7\,km\,s$^{-1}$ around each line.
For \prop{}, only the K=0 and K=1 transitions were considered, due to the low signal-to-noise ratio of the K=2 and K=3 transitions.
To enable comparison between the molecules in our following analysis, all maps were convolved to an angular resolution of 31", which corresponds to the half-power beam width (HPBW) of the largest observed beam.

\begin{table*}[h]
    \centering
    \caption{Spectroscopic parameters of the observed lines. }
    \begin{tabular}{l c c c c c c c c c}
    \hline\hline 
    \noalign{\smallskip}
    Molecule & Transition & Frequency\tablefootmark{a} & $F_\mathrm{eff}/B_\mathrm{eff}$ & rms & $E_\mathrm{up}$\tablefootmark{a} & $g_\mathrm{up}$\tablefootmark{a} & $A$\tablefootmark{a} & Reference \\
     & ($J'_{K'}-J_K$) & (MHz) &  & (mK) & (K) &  & (s$^{-1}$) & \\
    \noalign{\smallskip}
    \hline
    \noalign{\smallskip}
    HC$_3$N & $J=11-10$ & 100076.39(2) & 0.95/0.79 & 110 & 28.8 & 23 & 7.77$\times10^{-5}$ & 1 \\
    HCC$^{13}$CN$^*$ & $J=10-9$ & 90601.78(3) & 0.95/0.80 & 8 & 23.9 & 21 & 5.74$\times10^{-5}$ & 2 \\
    DC$_3$N$^*$ & $J=10-9$ & 84429.814(3) & 0.95/0.81 & 17 & 22.3 & 21 & 4.69$\times10^{-5}$ & 1\\
    DC$_3$N & $J=11-10$ & 92872.375(3) & 0.95/0.80 & 19 & 26.7 & 23 & 6.24$\times10^{-5}$ & 1\\
    \noalign{\smallskip}
    \hline
    \noalign{\smallskip}
    CH$_3$CCH & $J_K=5_3-4_3$      & 85442.6012(1) & 0.95/0.81 & 16 & 77.3 & 44 & $1.30\times10^{-6}$ & 3 \\
    CH$_3$CCH & $J_K=5_2-4_2$      & 85450.7663(1) & 0.95/0.81 & 16 & 41.1 & 22 & $1.70\times10^{-6}$ & 3  \\
    CH$_3$CCH$^*$ & $J_K=5_1-4_1$  & 85455.6667(1) & 0.95/0.81 & 10 & 19.5 & 22 & 1.95$\times10^{-6}$ & 3 \\
    CH$_3$CCH & $J_K=5_0-4_0$      & 85457.3003(1) & 0.95/0.81 & 10 & 12.3 & 22 & 2.03$\times10^{-6}$ & 3  \\
    CH$_3$CCH & $J_K=6_{3}-5_{3}$  & 102530.3476(1) & 0.95/0.79 & 37 & 82.3 & 52 & $2.69\times10^{-6}$ & 3  \\
    CH$_3$CCH & $J_K=6_{2}-5_{2}$  & 102540.1446(1) & 0.95/0.79 & 37 & 46.1 & 26 & $3.16\times10^{-6}$ & 3  \\
    CH$_3$CCH & $J_K=6_{1}-5_{1}$  & 102546.0242(1) & 0.95/0.79 & 37 & 24.5 & 26 & 3.46$\times10^{-6}$ & 3  \\
    CH$_3$CCH & $J_K=6_{0}-5_{0}$  & 102547.9844(1) & 0.95/0.79 & 33 & 17.2 & 26 & 3.56$\times10^{-6}$ & 3  \\
    CH$_3$CCD & $J_K=6_1-5_1$ & 93454.331(3) & 0.95/0.80 & 8 & 22.9 & 26 & 2.51$\times10^{-6}$ & 4 \\
    CH$_3$CCD$^*$ & $J_K=6_0-5_0$ & 93456.044(3) & 0.95/0.80 & 8 & 15.7 & 26 & 2.59$\times10^{-6}$ & 4 \\
    CH$_2$DCCH$^*$ & $J_{K_aK_c}=6_{06}-5_{05}$ & 97080.728(6) & 0.95/0.80 & 15 & 16.3 & 13 & 3.03$\times10^{-6}$ & 4 \\
    \noalign{\smallskip}
    \hline
    \end{tabular}
    \label{Tab:ObservedLinesAll}
    \tablefoot{The transitions used to derive the molecular column density are marked with an asterisk. Numbers in parenthesis give the uncertainty on the last digit.
    \tablefoottext{a}{Extracted from the Cologne Database for Molecular Spectroscopy \citep{Mueller2001}.}}
    \tablebib{
    (1) \cite{HC3Nref}, (2) \cite{H13C3Nref}, (3) \cite{CH3CCHref}, (4) \cite{CH3CCD_CH2DCCHref}.
    }
\end{table*}

\section{Analysis}\label{sec:analysis}

\subsection{Spectral line emission}

The goal of this work is to study the deuterium fraction of \cyano{} and \prop{} across the source. To do so, a reliable estimation of the molecular column densities is crucial.

\begin{figure*}[h]
    \centering
    \includegraphics[width=\hsize]{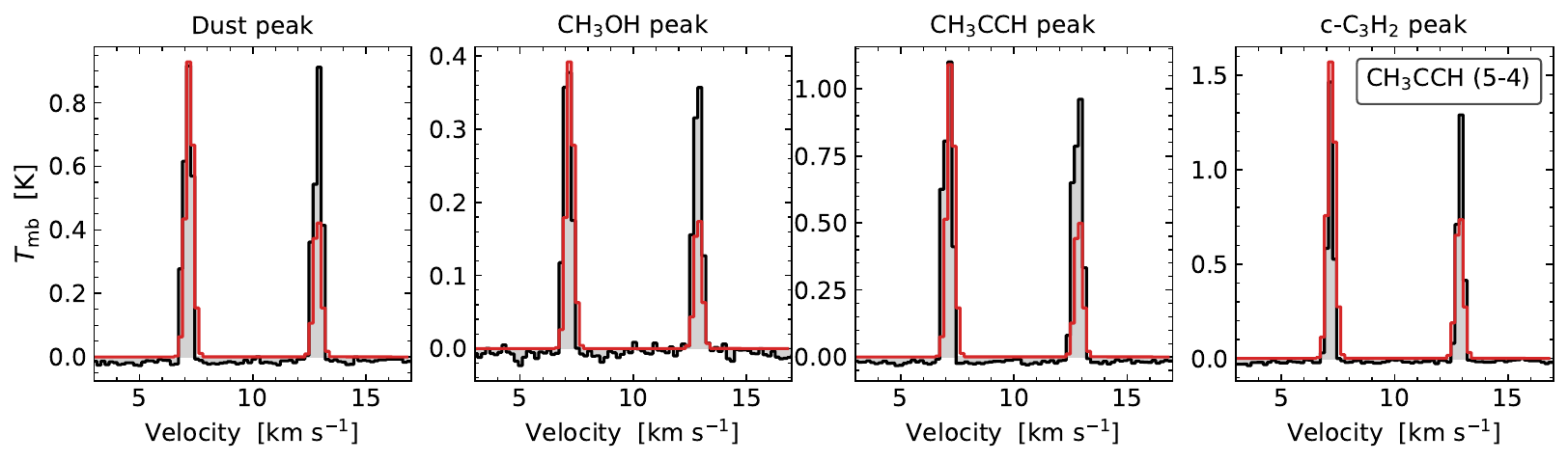}
    \includegraphics[width=\hsize]{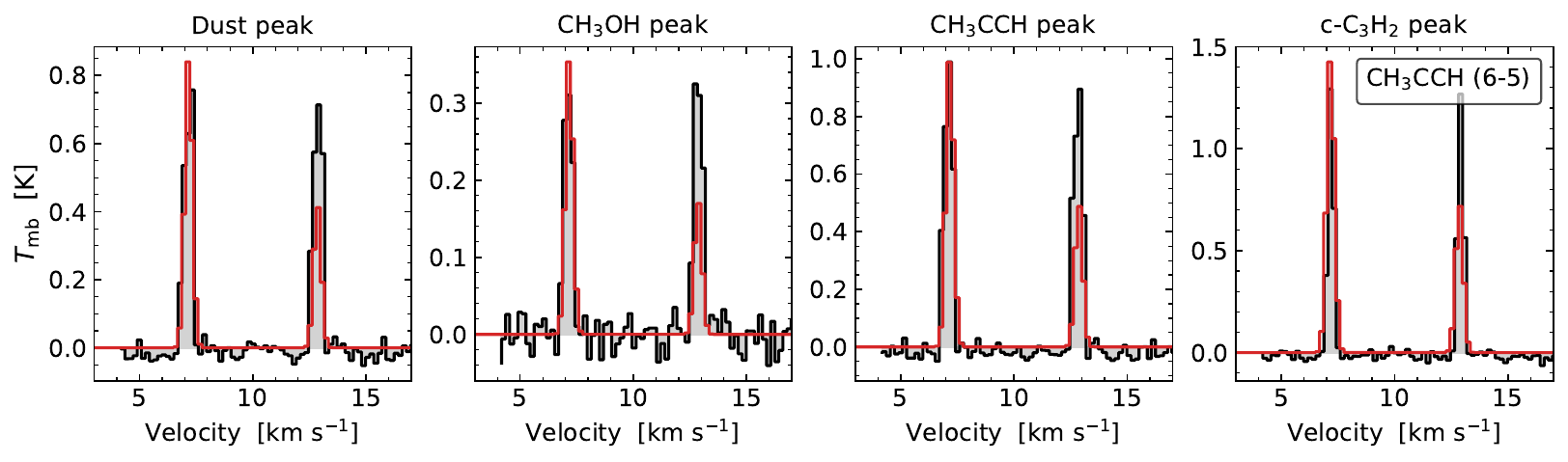}
    \includegraphics[width=\hsize]{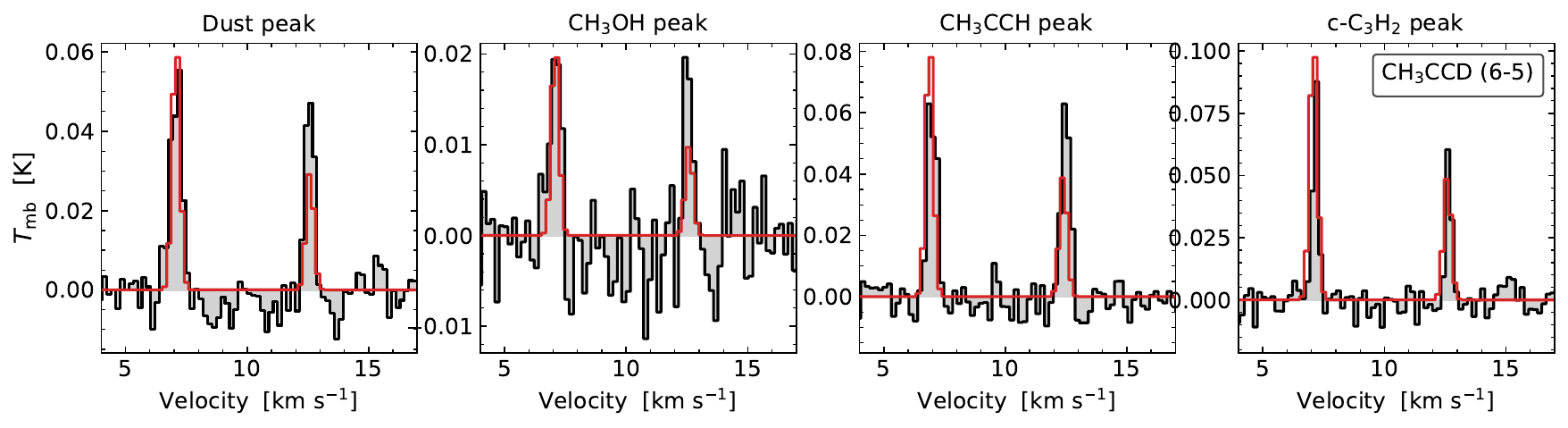}
    \caption{Observed spectra (black) of the \prop{} (\textit{top}) and the \propd{} (\textit{bottom}) K=0 and K=1 transitions extracted towards the three molecular peaks in L1544 and the dust peak, using a circular aperture with diameter 31". The extraction locations are indicated in Fig.~\ref{fig:IntegratedIntensityMaps}. Shown in red are synthetic spectra produced with the LTE model generator of the python package \texttt{pyspeckit}. The input column densities are [4, 2, 5, 8]$(\pm0.3)\times10^{13}$\,cm$^{-2}$ for \prop{} and [3, 1, 4, 5]$(\pm0.3)\times10^{12}$\,cm$^{-2}$ for \propd{} at the dust peak, \meth{} peak, \prop{} peak, and \cyc{} peak, respectively. The input excitation temperature is set to a constant value of 10\,K for both molecules.}
    \label{fig:CH3CCHmoleculepeaks}
\end{figure*}

For \prop{} and \propd{}, we observe more than one transition.  
Figure~\ref{fig:CH3CCHmoleculepeaks} presents spectra of the \prop{} (5--4) and (6--5), and the \propd{} (6--5) K=0 and K=1 transitions, extracted towards the three molecular peaks and the dust peak in L1544 by using a circular aperture with a diameter corresponding to the telescope beam size, 31". The extraction locations and beam size are indicated in Fig.~\ref{fig:IntegratedIntensityMaps}.
The observations show that the line intensity ratio between the K=0 and K=1 transitions varies across the core, for \prop{} between 1:1 and 1:1.2, and for \propd{} between 1:1 and 1:1.4.
In a first approach, we assumed local thermodynamic equilibrium (LTE) and modelled the spectra using \texttt{lte\_molecule} and \texttt{generate\_model} from the python package \texttt{pyspeckit}.
We applied a constant excitation temperature of 10\,K, which is consistent with previous measurements of \prop{} in this source \citep[e.g.][]{Vastel2014}, and adjusted the column densities to fit the intensity of the K=0 lines.
The resulting modelled spectra are shown in red in Fig.~\ref{fig:CH3CCHmoleculepeaks}.
For both \prop{} and \propd{}, the line ratio between K=0 and K=1 is estimated to be around 2:1.
This is not visible in the observations, which vary between 1:1 and 1:1.4.
However, the emission lines are in the optically thin regime, with $\tau_\mathrm{max}(T_\mathrm{ex}=10\,\mathrm{K})\approx0.27$. 
Therefore, self-absorption is unlikely to be the cause of the line ratios. 
Following this, there must be another effect influencing the line intensities, and given the relatively high critical density  of \prop{} ($n_\mathrm{crit}\approx10^5$\,cm$^{-3}$) compared to the densities across the core ($\rm 10^4-10^7\,cm^{-3}$), it is likely that non-LTE effects play a role.

Therefore, a non-LTE approach would be necessary to test whether non-LTE effects might cause the observed intensity ratios, and to constrain the molecular column densities. 
A full non-LTE analysis lies beyond the scope of this paper, because the required 3D model of the temperature and density structure of the core does not exist. 
Instead, we focus on the analysis of the emission lines extracted at the dust peak of L1544.
In the following, we model the observed spectra with non-LTE radiative transfer in Sec.~\ref{sec:LOCmodelling}, before we analyse the column densities and deuterium fractions in LTE in Sec.~\ref{sec:ColDensandDeutMaps}, using the non-LTE results to constrain the physical parameters of the emission.

\subsection{Non-LTE modelling at the dust peak}\label{sec:LOCmodelling}

\begin{figure}[h]
    \centering
    \includegraphics[width=\hsize]{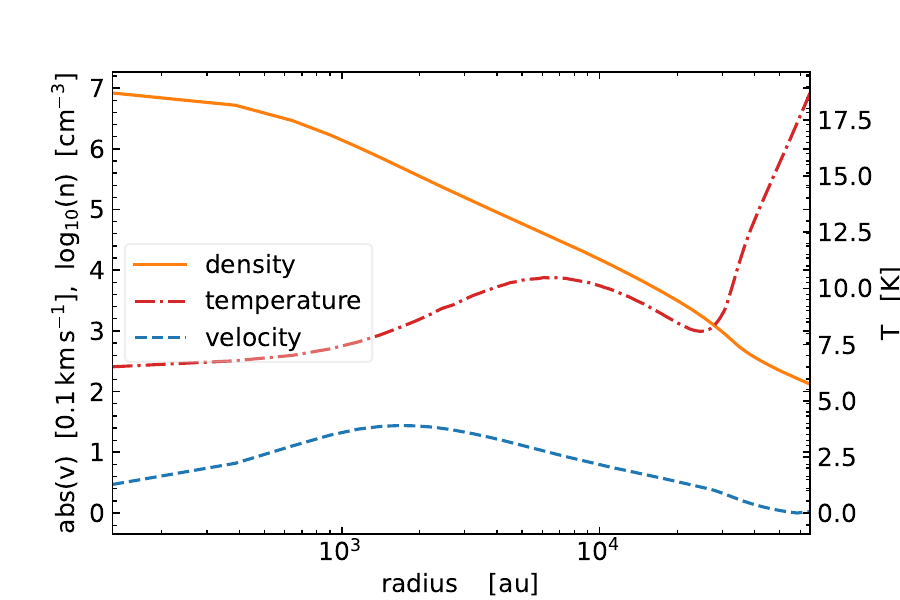}
    \caption{Profiles of the gas temperature (red), H$_2$ number density (orange, in logarithmic scale), and infall velocity (blue, in units of 0.1\,km\,s$^{-1}$) for the Keto-Caselli model of L1544 \citep{Keto2015}. The velocity in the model is negative but is shown here as positive to improve readability.}
    \label{fig:KetoCasellimodel}
\end{figure}

In this section, we simulate the observed molecular spectra towards the dust peak of L1544 with the non-LTE radiative transfer code Line Transfer with OpenCL (LOC, \citealt{Juvela2020}). 
We assumed a 1D model that considers spherical symmetry of the physical properties and chemical abundances.
The physical structures are characterised by volume density, kinetic temperature, and infall velocity.
For this, we adopted the physical model of L1544 presented in \cite{Keto2015} (hereafter Keto-Caselli model), shown in Fig.~\ref{fig:KetoCasellimodel}. 
It describes an unstable quasi-equilibrium Bonnor-Ebert sphere \citep{Ebert1955,Bonnor1956} with a peak central H$_2$ volume density of $n_0\approx10^7$\,cm$^{-3}$ and a central gas temperature of 6\,K. 
To approximate the molecular abundances, we used profiles of the fractional abundance with respect to H$_2$ molecules, hereafter referred to as 'abundance profiles'. We applied both constant abundance profiles, as well as radial abundance profiles predicted by the state-of-the-art gas-grain chemical model \texttt{pyRate} \citep{Sipila2019}.
For the constant abundance profiles, we set the abundance to zero within a radius of 1800\,au, as L1544 is known to exhibit a freeze-out of 99.99\% of species heavier than Helium in this central region \citep{Caselli2022}.
In pyRate, we applied the two-phase model of the code, where the gas-phase chemistry and the entire ice chemistry on the grains are active.
We used the same initial abundances and chemical networks as presented in \cite{Sipila2019} and the same standard values for various model parameters as used in \cite{Giers2022}. 
In the chemical simulation, the Keto-Caselli model is used as a static physical model for L1544, assuming an external visual extinction of $A_V=2$\,mag to account for the molecular cloud where the core is embedded.
The resulting abundance profiles were extracted at various evolutionary times across the chemical simulation.

For the line simulations of \cyano{} and \prop{}, the latest collision rate coefficients are available on the Leiden Atomic and Molecular Database (LAMDA, \citealt{Schoier2005,Faure2016,BenKhalifa2024}). 
In the case of \xcyano{}, we used the rates of \cyano{} and scaled them with the corresponding reduced mass of the isotopologue. 
For the deuterated isotopologues of \cyano{} and \prop{}, a scaling of the rates was not possible, as the different nuclear spins induce different hyperfine structures. Therefore, we could not apply the non-LTE modelling to these species.

\begin{figure}[h]
    \centering
    \includegraphics[width=\hsize]{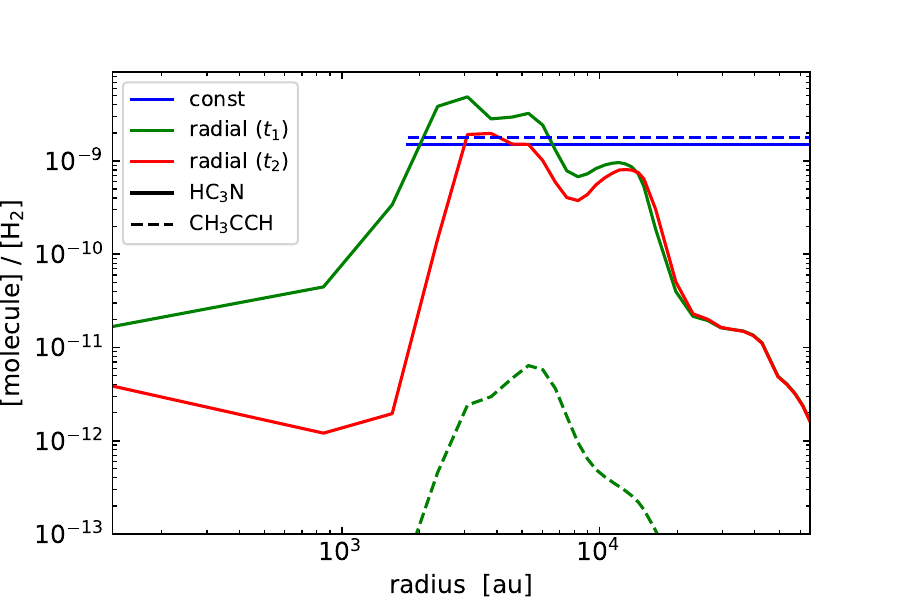}
    \caption{Fractional abundance profiles of \cyano{} (solid) and \prop{} (dashed) of the best-fit results produced with LOC. For \xcyano{}, the radial abundance profiles derived with chemical modelling (green, red) correspond to the profiles of the main isotopologue, scaled down by the isotopic ratio, 68. }
    \label{fig:HC3Nabundances}
\end{figure}

\begin{figure}[h]
    \centering
     \includegraphics[width=\hsize]{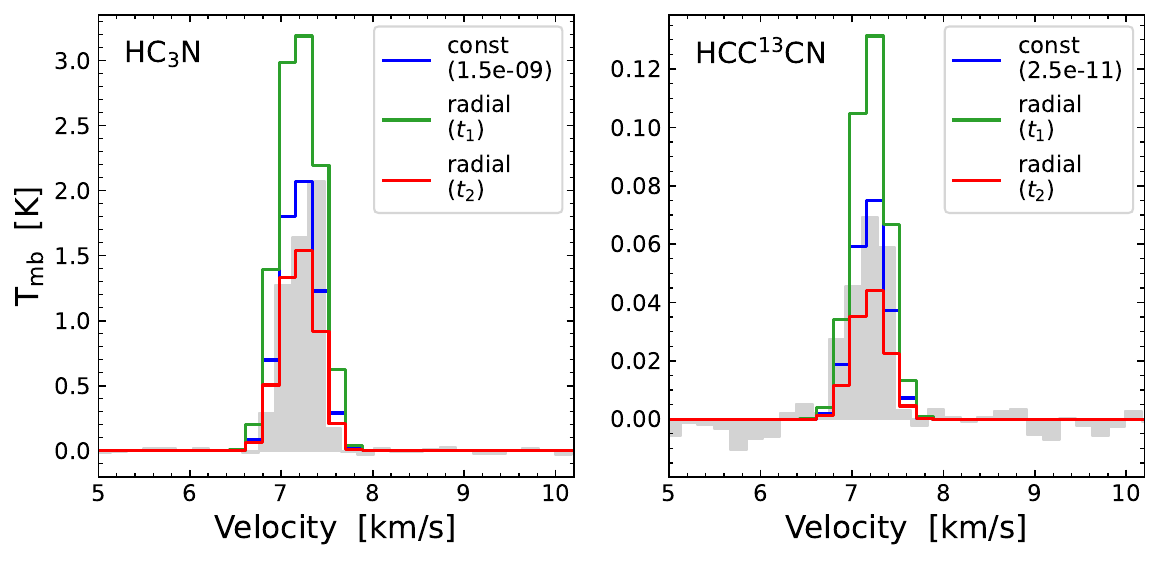}
    \caption{Comparison of the observed spectra (grey) of \cyano{}, and \xcyano{}, extracted towards the dust peak of L1544, to the synthetic spectra (red, blue, green) produced with LOC. For the constant abundances, a depletion radius of 1800\,au is assumed.}
    \label{fig:LOCHC3N}
\end{figure}

\begin{figure*}[h]
    \centering
    \includegraphics[width=\textwidth]{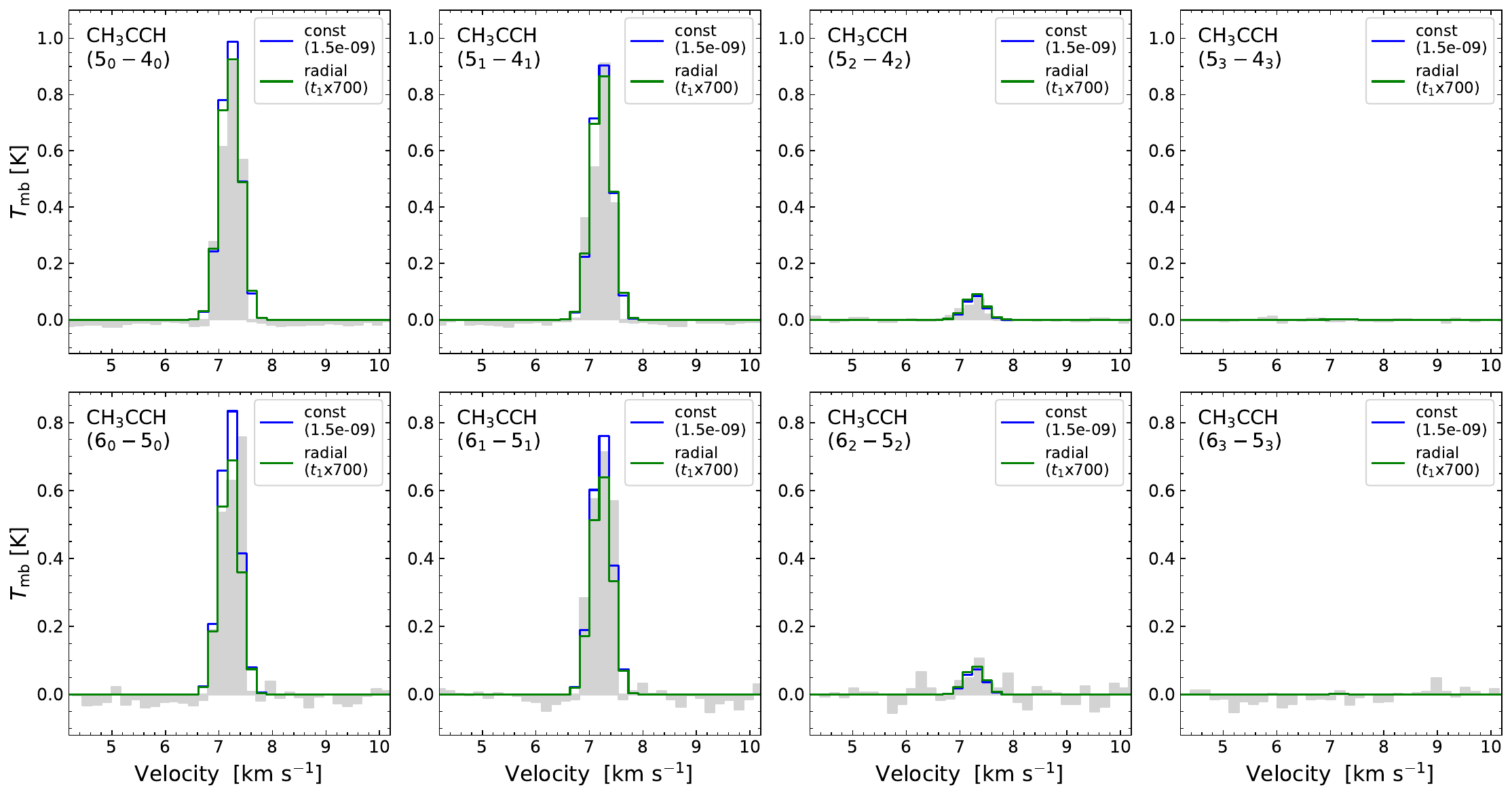}
    \caption{Comparison of the observed spectra (grey) of \prop{}, extracted towards the dust peak of L1544, to the synthetic spectra (red, blue, green) produced with LOC. For the constant abundances, a depletion radius of 1800\,au is assumed. }
    \label{fig:LOCCH3CCH}
\end{figure*}

The abundance profiles that provide the best-fit solutions for \cyano{} and \prop{} are shown in Fig.~\ref{fig:HC3Nabundances}.
A comparison between the modelled spectra and the observed lines is shown in Fig.~\ref{fig:LOCHC3N} for \cyano{}, and \xcyano{}, and in Fig.~\ref{fig:LOCCH3CCH} for \prop{}. The observed spectra, shown in black, are extracted at the dust peak of the core, averaged over an area corresponding to the telescope beam size (31").
To enable a comparison, the synthetic spectra were convolved with this observational beam for each transition frequency.
The molecular column densities at the dust peak derived by LOC are given in Table~\ref{Tab:LOCresultsTexNcol}.

The spectrum of \xcyano{} can be reproduced both with a constant abundance and a radial abundance profile from the chemical modelling. With a depletion radius of 1800\,au, a constant abundance of $2.5\cdot10^{-11}$ fits best.
When applying the abundance profiles from pyRate, the observed lines are reproduced within a factor of 2 with profiles extracted at early time steps of the chemical simulation ($t_1=10^5$\,yrs, $t_2=1.3\cdot10^5$\,yrs).

The brightness of the \cyano{} spectral line can be reproduced with modelling at a constant fractional abundance of $1.5\cdot10^{-9}$ with a depletion radius of 1800\,au, and, similar to \xcyano{}, at early time steps within a factor of 2 in intensity ($t_1=10^5$\,yrs, $t_2=1.3\cdot10^5$\,yrs). However, the modelling fails to recreate the slightly asymmetric, redshifted shape of the line profile. 
This asymmetric shape is likely caused by the velocity gradient across the core with higher velocities in the south, where \cyano{} peaks \citep{Bianchi2023}. Therefore, more material emits at slightly higher velocities, compared to the systemic velocity of the core in the centre, causing a redshifted asymmetry.
The column densities of \cyano{} and \xcyano{} derived by the best-fit models result in an isotopic ratio of $^{12}$C/$^{13}$C$=60\pm20$, which is in agreement with the value for the local interstellar medium, $^{12}$C/$^{13}$C$=68$ \citep{Milam2005}.

All observed transitions of \prop{} can be reproduced with a constant abundance profile of $1.5\cdot10^{-9}$ and a depletion radius of 1800\,au. 
As the non-LTE modelling is able to reproduce the observed line ratios between the K=0 and K=1 transitions correctly, we are able to use the central column density of \prop{} derived by LOC in the analysis below. 
However, the emission lines cannot be reproduced with any of the abundance profiles derived from chemical modelling. 
In fact, the radially varying abundance profiles largely underestimate the abundance of \prop{}, by roughly three orders of magnitude.
Figure~\ref{fig:LOCCH3CCH} illustrates that it is necessary to multiply the abundance profile by a factor of 700 to reproduce the observed line intensities.
This issue was already noted in previous research, where various chemical models failed to predict observed molecular abundances of \prop{}. 
The formation pathways of \prop{} have been widely studied, with formation proposed via ion-molecule reactions, neutral-neutral reactions, and dissociative recombination in the gas phase \citep{SchiffBohme1979,Turner1999,Calcutt2019}, and hydrogenation of C$_3$ on grain surfaces \citep{Hickson2016,Guzman2018}.
However, models demonstrated that these gas-phase and grain-surface formation pathways are not sufficient to reproduce \prop{} abundances in cold molecular clouds \citep[see e.g.][]{Oberg2013,Hickson2016}.
Hence, it is necessary to further investigate potential grain surface reactions of \prop{} or alternative gas phase pathways at low temperatures to improve the chemical networks and ultimately the model predictions.

\begin{table}[h]
    \centering
    \caption{Molecular column densities for the best-fit models, derived with LOC towards the dust peak (see Figs.~\ref{fig:LOCHC3N} and \ref{fig:LOCCH3CCH}).}
    \begin{tabular}{l l c}
    \hline\hline 
    \noalign{\smallskip}
    Molecule & model & $N_\mathrm{centre}^\mathrm{NLTE}$ \\
             &       & ($\times10^{12}$\,cm$^{-2}$) \\
    \noalign{\smallskip}
    \hline
    \noalign{\smallskip}
    \cyano{}  & const & 38 \\
              & radial ($t_1$) & 64 \\
              & radial ($t_2$) & 21 \\
    \xcyano{} & const & 0.63 \\
              & radial ($t_1$) & 0.94 \\
              & radial ($t_2$) & 0.31 \\
    \prop{}   & const & 45 \\
    \noalign{\smallskip}
    \hline
    \end{tabular}
    \label{Tab:LOCresultsTexNcol}
\end{table}


\subsection{Column density and deuteration maps}\label{sec:ColDensandDeutMaps}

A detailed non-LTE analysis is only possible towards the dust peak of L1544. 
To study the distribution of the molecular emission across the core, we used the LTE assumptions to compute the column density maps, and used the non-LTE results as a constraint.

\subsubsection{Column density}
To derive the molecular column densities, we assumed optically thin emission, following the derivation presented in \cite{Mangum2015}. Furthermore, we applied the approximation of a constant excitation temperature across the core \citep[see][]{Caselli2002b,Redaelli2019}. This results in a column density of
\begin{equation}\label{equ:columndensity}
    N=\frac{8\pi\nu^3}{c^3}\frac{Q_\mathrm{rot}(T_\mathrm{ex})}{g_uA_\mathrm{ul}}\left[J_\nu(T_\mathrm{ex})-J_\nu(T_\mathrm{bg})\right]^{-1}\frac{\mathrm{e}^{\frac{E_u}{kT_\mathrm{ex}}}}{\mathrm{e}^{\frac{h\nu}{kT_\mathrm{ex}}}-1}\int T_\mathrm{mb}\mathrm{dv}\;,
\end{equation}
where $Q_\mathrm{rot}(T_\mathrm{ex})$ is the partition function of the molecule at an excitation temperature $T_\mathrm{ex}$, $g_u$ and $E_u$ respectively represent the degeneracy and energy of the upper level of the transition, $A_\mathrm{ul}$ is the Einstein coefficient for spontaneous emission, $T_\mathrm{bg}=2.73$\,K is the temperature of the cosmic microwave background, $J(T)$ is the Rayleigh-Jeans equivalent temperature, and $T_\mathrm{mb}$ is the main-beam temperature. 
The corresponding parameters for each transition used in the derivation of the column density are listed in Table \ref{Tab:ObservedLinesAll}.

The observed range of $E_u$ only covers about 5\,K for the molecules where multiple transitions were observed (\prop{}, \propd{}, \dcyano{}). Therefore, an excitation analysis to derive the excitation temperatures pixel by pixel is not feasible.
Instead, we followed \cite{Redaelli2019} to determine the excitation temperatures for \prop{} and \xcyano{} and fixed the column density at the dust peak to the value derived from a constant molecular abundance profile with the non-LTE modelling in Sect.~\ref{sec:LOCmodelling} (see Table~\ref{Tab:LOCresultsTexNcol}). From this, we derived the corresponding excitation temperatures in the LTE regime, using Eq.~\ref{equ:columndensity}, and applied this value across the whole core to derive the column density maps.

Therefore, we set an excitation temperature of 7\,K for \xcyano{} and 10\,K for \prop{}, which is in agreement with values used in previous studies \citep[see e.g.][]{Howe1994,Markwick2005,Vastel2014,HilyBlant2018b,Agundez2019,Bianchi2023}.
As shown in the previous section, non-LTE effects are responsible for the line intensity ratio observed for \prop{} K=0 and K=1 towards the dust peak. 
We decided to use the K=0 lines to derive the column density map for \prop{}, because the column density value of these lines at the dust peak matches with an excitation temperature that is reasonable for L1544. 
For the K=1 lines, however, we would have to use an excitation temperature around 60\,K, which is significantly higher than what was routinely observed in this core.
An example of the \prop{} column density and deuteration maps derived from the K=1 transitions is given in Appendix~\ref{sec:app:DeuterationOfCH3CCH}.
For the further analysis, we focus on the K=0 line of the (5--4) transition instead of the (6--5) transition because of its higher signal-to-noise ratio.

As the non-LTE approach was not possible for the deuterated isotopologues due to missing collision rate coefficients, we adopted the respective excitation temperature of the main species. 
To derive the column density of \propd{}, we used the K=0 transition, as for \prop{}.
The effect of the excitation temperature on the derived column density ratios was found to be weak, with a change of a few percent upon a variation of $\pm$1\,K \citep[as stated in ][]{Spezzano2013}.

To account for the varying optical depth of the lines across the core, we corrected the column density by deriving the optical depth at each pixel individually using the respective excitation temperature, then multiplying the maps by \citep[see e.g.][]{GoldsmithLanger1999}
\begin{equation}\label{equ:opticaldepthcorrection}
    f_\tau = \frac{\tau}{1-\mathrm{e}^{-\tau}}\;.
\end{equation}
This results in a correction factor of up to 1.1 in the southern part of the core.

\begin{figure*}
    \includegraphics[width=0.33\textwidth]{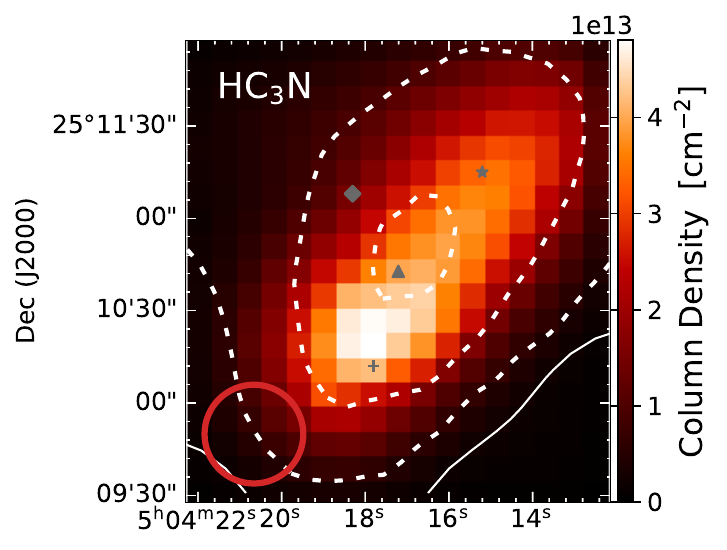}
    \includegraphics[width=0.33\textwidth]{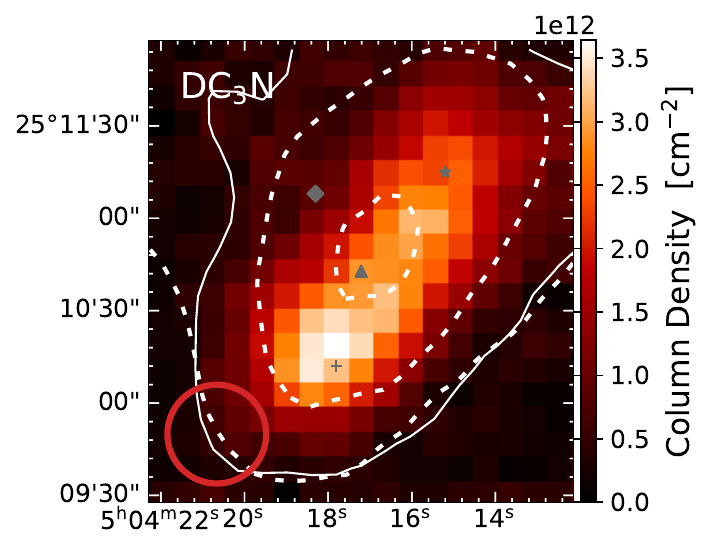}
    \includegraphics[width=0.32\textwidth]{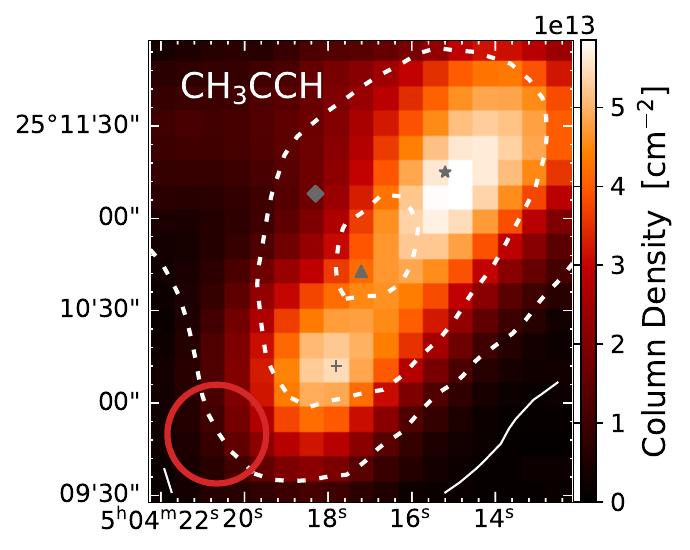}
    \includegraphics[width=0.34\textwidth]{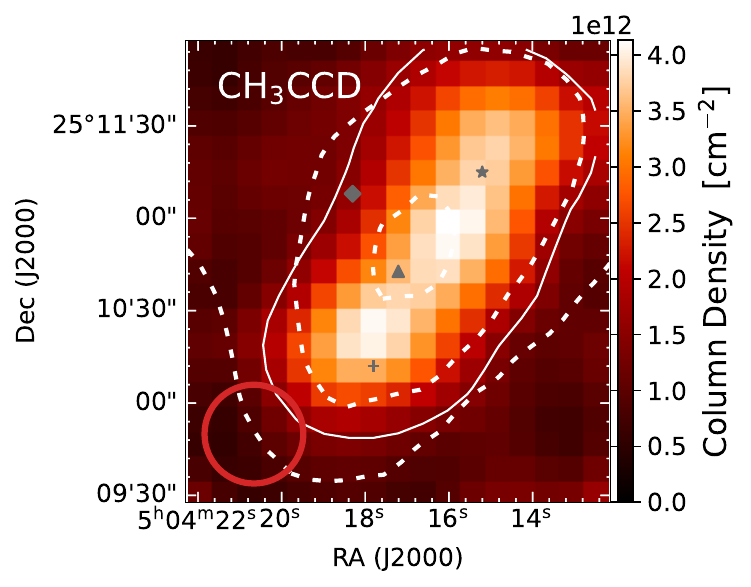}
    \includegraphics[width=0.33\textwidth]{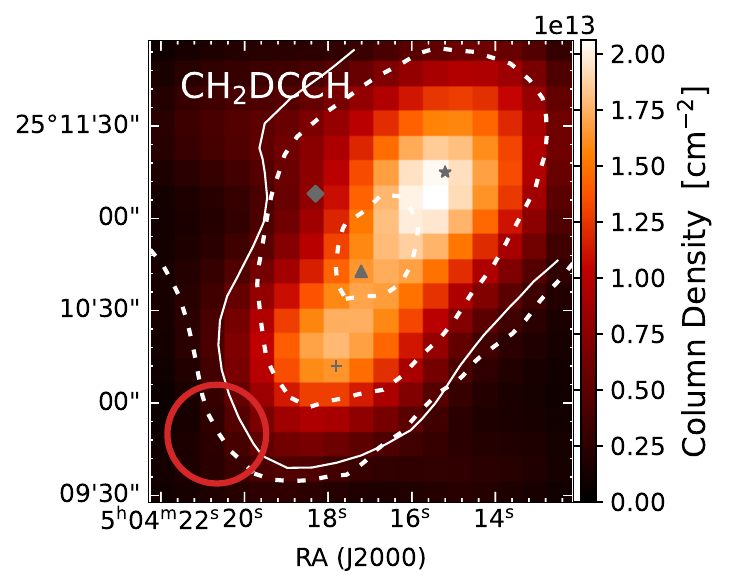}
    \caption{Column density maps of \cyano\, and \prop{} and their deuterated isotopologues. The column densities are derived with a constant excitation temperature of 7\,K and 10\,K for \cyano{} and \prop{} and their isotopologues, respectively, and are corrected for optical depth. 
    To derive the column density of \cyano{}, we used the emission of \xcyano{}, assuming a constant $^{12}$C/$^{13}$C=68 \citep{Milam2005}. 
    The solid line contours indicate the 3$\sigma$ level of the integrated intensity. The dashed line contours represent 30\%, 50\%, and 90\% of the H$_2$ column density peak derived from $Herschel$ maps \citep{Spezzano2016}. The circle in the bottom-left corner indicates the beam size of the IRAM 30\,m telescope (31"). The markers in grey represent the dust peak (triangle) and the molecular emission peaks of \meth{} (diamond), \prop{} (star), and \cyc{} (plus sign).}
    \label{fig:columndensitymaps}
\end{figure*}

The transitions used to derive the column density for each species are marked with an asterisk in Table~\ref{Tab:ObservedLinesAll}.
The corresponding column density maps are presented in Fig.~\ref{fig:columndensitymaps}.
In the following, we discuss the results for the each species individually.

\paragraph{\cyano{}:}
The emission of \cyano{} is moderately thick in this core, with optical depths of up to 0.9. Therefore, we approximated the column density of \cyano{} with the emission of the $^{13}$C isotopologue, multiplied by the isotopic ratio for the local interstellar medium, $^{12}\text{C}/^{13}\text{C}=68$. 
Given the results of the non-LTE modelling, $^{12}\text{C}/^{13}\text{C}=60\pm20$ at the dust peak, this is a reasonable assumption.
The resulting column density map peaks in the south-east of the core (see Fig.~\ref{fig:columndensitymaps}), on the carbon-chain peak caused by non-uniform external illumination \citep[see ][]{Spezzano2017}, with a maximum value of 6.3(5)$\times10^{13}$\,cm$^{-2}$, where the number in parenthesis indicates the uncertainty of the last digit. This is consistent with previous measurements towards L1544 (2-8$\times10^{13}$\,cm$^{-2}$, \citealt{HilyBlant2018b}; 7(2)$\times10^{13}$\,cm$^{-2}$, \citealt{Howe1994}).
Additionally, the column density of \cyano{} shows a local peak towards the north-west of the core (where \prop{} peaks) with a maximum of 5.4(5)$\times10^{13}$\,cm$^{-2}$, and a dip in the core centre indicating the molecular freeze-out zone.
The deuterated isotopologue is distributed similarly to the $^{13}$C-isotopologue, peaking in the south-east (3.5(2)$\times10^{12}$\,cm$^{-2}$), and with a local peak in the north-west (3.1(2)$\times10^{12}$\,cm$^{-2}$). These column density levels are consistent with previous measurements towards L1544 \citep[e.g.][]{Howe1994}.

\paragraph{\prop{}:}
In contrast to \cyano{}, the emission of \prop{} peaks in the north-west of the core, with a maximum value of 5.85(6)$\times10^{13}$\,cm$^{-2}$. Here, it is possible that ongoing accretion of material from cloud to core leads to chemically fresh gas and a replenishment of \prop{} \citep[see][]{Giers2025}. A weaker, secondary peak is visible in the south-east of the core, on the carbon-chain peak, with values of 5.43(6)$\times10^{13}$\,cm$^{-2}$.
The column density of \dprop{} is distributed similarly to the main isotopologue, with a peak column density of 2.0(2)$\times10^{13}$\,cm$^{-2}$ and a southern peak at 1.7(2)$\times10^{13}$\,cm$^{-2}$, which is equal to the northern peak within error bars. 
The column density of \propd{}, on the other hand, peaks closer to the core centre (4.1(4)$\times10^{12}$\,cm$^{-2}$), with a second, equally strong peak (4.0(5)$\times10^{12}$\,cm$^{-2}$) in the southern part of the core.

\subsubsection{Deuterium fraction}

\begin{figure}[h]
    \centering
    \includegraphics[width=\hsize]{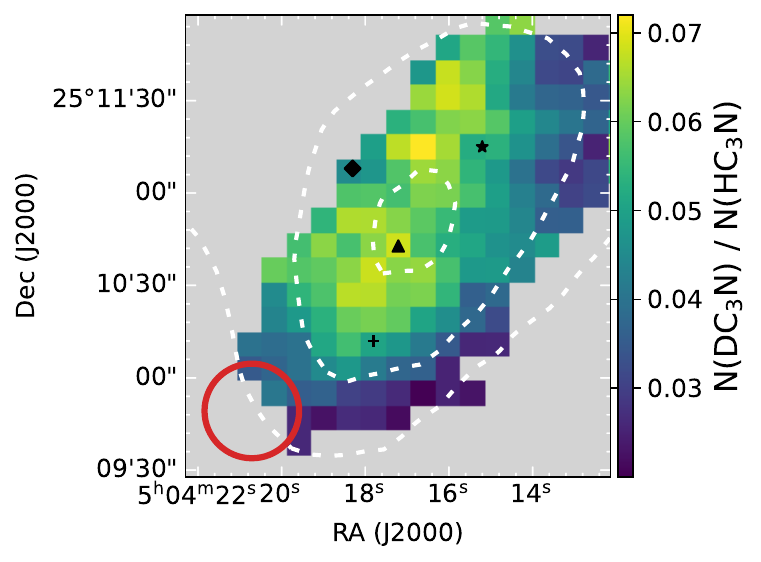}
    \caption{Deuteration map of \cyano{}. Only pixels above the 3$\sigma$ level of the respective integrated intensities are plotted. 
    The dashed line contours represent 30\%, 50\%, and 90\% of the H$_2$ column density peak derived from $Herschel$ maps \citep{Spezzano2016}. The circle in the bottom-left corner indicates the beam size of the IRAM 30\,m telescope (31"). The markers in black represent the dust peak (triangle) and the molecular emission peaks of \meth{} (diamond), \prop{} (star), and \cyc{} (plus sign).
    The corresponding D/H error map is shown in Fig.~\ref{fig:DeuterationErrormaps}.}
    \label{fig:deuterationmapsHC3N}
\end{figure}

\begin{figure*}[h]
    \centering
    \includegraphics[width=0.34\textwidth]{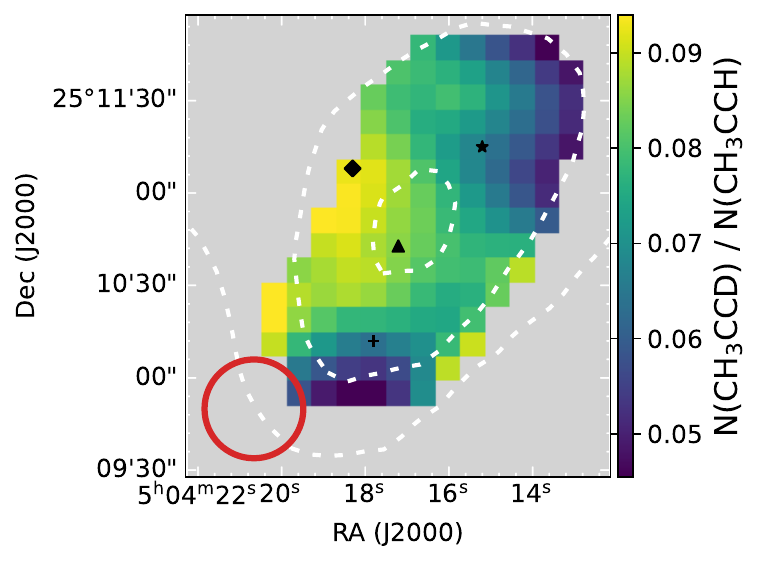}
    \includegraphics[width=0.32\textwidth]{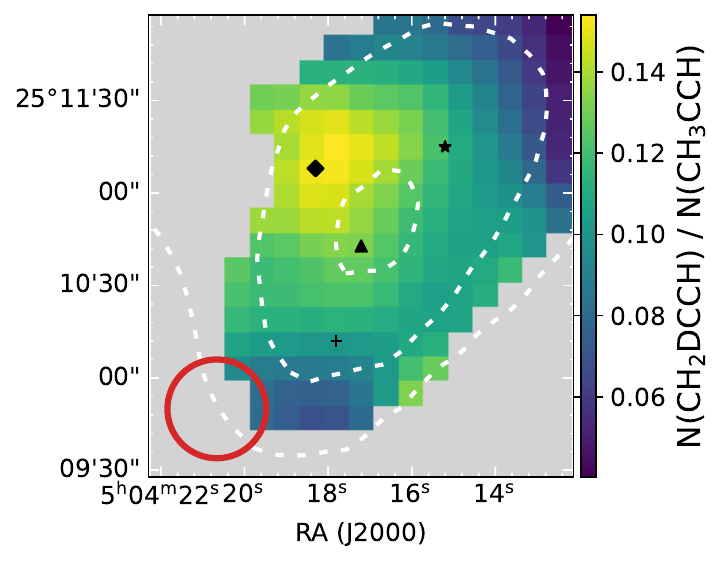}
    \includegraphics[width=0.32\textwidth]{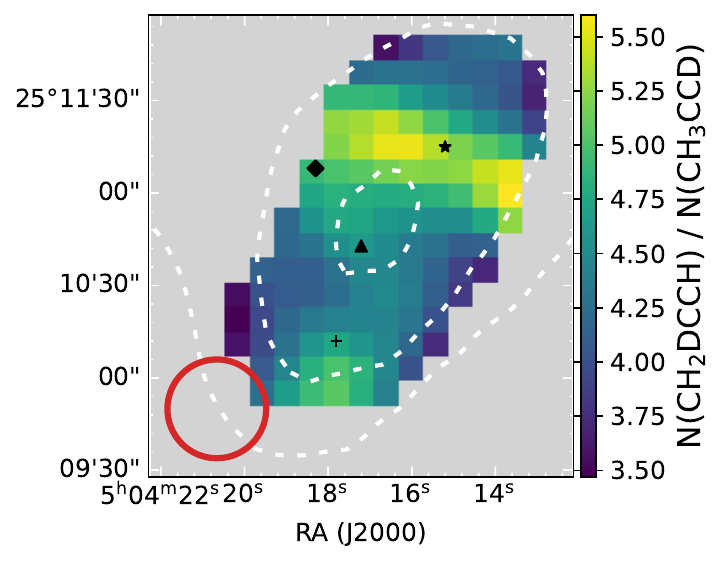}
    \caption{Deuteration maps of \prop{}. Only pixels above the 3$\sigma$ level of the respective integrated intensities are plotted. 
    The dashed line contours represent 30\%, 50\%, and 90\% of the H$_2$ column density peak derived from $Herschel$ maps \citep{Spezzano2016}. The circle in the bottom-left corner indicates the beam size of the IRAM 30\,m telescope (31"). The markers in black represent the dust peak (triangle) and the molecular emission peaks of \meth{} (diamond), \prop{} (star), and \cyc{} (plus sign). The corresponding D/H error maps are shown in Fig.~\ref{fig:DeuterationErrormaps}.}
    \label{fig:deuterationmapsCH3CCH}
\end{figure*}

The deuterium fractions were derived pixel by pixel by dividing the column density maps of the deuterated isotopologues by the column density map of the corresponding main species. 
For the ratio of N(\dprop{})/N(\prop{}), we accounted for the degeneracy of the D atom by dividing the ratio by a factor of 3.
The resulting deuteration maps are presented in Fig.~\ref{fig:deuterationmapsHC3N} for \cyano{}, and Fig.~\ref{fig:deuterationmapsCH3CCH} for \prop{}.
In the following, we describe the results for each molecule individually:

\paragraph{\dcyano{}/\cyano{}:}
The deuterium fraction map shows moderate levels of deuteration, in the range of 0.04 - 0.07. The distribution is rather homogeneous across the whole core. 
The highest values (0.06-0.07) are widespread, and located mostly north and south-east of the core centre.

\paragraph{\propd{}/\prop{}:}
The map shows that the deuterium fraction is, within the uncertainties (see Fig.~\ref{fig:DeuterationErrormaps}), 
distributed rather homogeneously across the core, with values between 0.07 and 0.09. 
A trend of higher values ($\geq$0.09) is visible towards the centre and the north-east of the core. 
The high D/H values at the edges of the map should be treated with caution since they have large uncertainties of around 30\%.

\paragraph{\dprop{}/\prop{}:}
This deuteration map shows a larger extension across the core than the map of \propd{}/\prop{}, due to a better signal-to-noise ratio.
Additionally, the map exhibits a significantly higher deuterium fraction than \propd{}/\prop{}, with values between 0.09 and 0.15. 
In contrast to \propd{}/\prop{}, the distribution of \dprop{}/\prop{} exhibits a clear peak away from the centre towards the north-east of the core, with a peak value of 0.15(3). 

\paragraph{\dprop{}/\propd{}:}
The ratio map between the two deuterated isotopologues shows that \dprop{} is more abundant than \propd{}. 
The distribution is roughly homogeneous across the core, with a slight trend of higher values towards the north ($5.3\pm0.9$) compared to the south ($4.1\pm0.9$).
The statistically expected ratio of \dprop{}/\propd{} is 3, which results from the statistical considerations for the insertion of deuterium in the CH$_3$ group and the CCH group. 
We derive an average ratio of 4.5(8), which is a 50\% increase with respect to the statistical value.

\section{Discussion}\label{sec:discussion}

\subsection{\cyano{}}
The column density maps presented in Fig.~\ref{fig:columndensitymaps} show that the two carbon chains \cyano{} and \prop{} and their deuterated counterparts exhibit different morphologies across the core.
The emission of both \cyano{} and its deuterated isotopologue peak at the carbon-chain peak in the south of the core. 
The resulting deuterium fraction map, however, exhibits a rather homogeneous distribution across the core with moderate levels of deuteration (0.05-0.07).
This is in agreement with previous measurements in L1544 \citep{Howe1994}. 
Results for the deuterium fraction of \cyano{} show roughly similar values in other cores: 
0.03-0.11 in the dark cloud TMC-1 \citep{Howe1994},
0.03(1) in the cold envelope of the Class 0 protostar L1527 \citep{Sakai2009},
and 0.03(1) in the dense core L483 \citep{Turner2001,Sakai2009,Agundez2019}.

The moderate deuteration levels and extended morphology of the deuteration map in L1544 indicate that \cyano{} and \dcyano{} do not trace the high-density regions in the core centre (e.g. $\rm N_2H^+$ and $\rm N_2D^+$), but rather a less dense layer of gas in an outer shell of the core \citep[in agreement with][]{Rivilla2020}, at densities of less than $\rm 10^5\,cm^{-3}$ (see Fig.~\ref{fig:HC3Nabundances}).
In the less dense gas, the process of deuteration is less efficient due to a lower abundance of $\rm H_2D^+$.
Therefore, the moderate deuterium fraction suggests that \dcyano{} is most likely formed in the gas phase, presumably via ion-molecule reactions, as discussed in \cite{Rivilla2020}.

The assumption that \cyano{} and \dcyano{} trace intermediate-density gas in L1544 is supported by the abundance profiles used to reproduce the emission lines of \cyano{} and \xcyano{} (see Fig.~\ref{fig:HC3Nabundances}).
These abundance profiles peak at a volume density of roughly $10^5$\,cm$^{-3}$, which is significantly lower than the critical densities of both transitions ($n_\mathrm{crit}($\cyano{}$)\approx10^7$\,cm$^{-3}$, $n_\mathrm{crit}($\xcyano{}$)\approx10^6$\,cm$^{-3}$). 
This was also observed by \cite{Bianchi2023}, who state that the carbon-chain abundance is enhanced towards the external part of the core, where the material is more exposed to the ISRF (see also \citealt{Spezzano2016}).
This becomes visible in the slightly asymmetric line shapes of \cyano{} and \xcyano{} at the dust peak and the \cyc{} peak (see Fig.~\ref{fig:peakspectra}). 
This asymmetry is caused by the velocity gradient across L1544, which leads to higher velocities at the carbon-chain peak in the southern part \citep[see][]{Spezzano2016,Spezzano2017,Giers2025}, creating redshifted, asymmetric line profiles for carbon-chain molecules, as most of their emission originates in this part.
This effect cannot be reproduced with non-LTE modelling, because the physical structure assumes the core to be spherically symmetric and does not account for asymmetric velocity variations across the core.

\subsection{\prop{}}

The observed ratios between the peak intensities of the K=0 and K=1 emission lines of \prop{} and \propd{} -- both for the (5--4) and the (6--5) transitions -- cannot be reproduced assuming LTE conditions but instead require a non-LTE treatment. 
Compared with the observations, the LTE model either overestimates the K=0 line or underestimates the K=1 line.
Low densities in the layers traced by \prop{} could cause non-LTE conditions, leading to a sub-thermal excitation of the transitions and subsequently generating the emission lines different from the expected LTE profile.
Figure~\ref{fig:CH3CCHmoleculepeaks} shows that the line intensity ratios also vary across the core. 
However, the ratio of the integrated intensity maps of the ($5_0-4_0$) over the ($5_1-4_1$) transition is roughly constant across the core, with an average value of 1.05(2). 
Therefore, the effect of the line variations on the derived column density of \prop{} and the deuteration maps should be within the error bars constrained by the derivation.
The emission of \prop{} peaks in the north-west of L1544, away from the carbon-chain peak in the south-east.
The northern part of this core is shaped by the interaction of two filaments \citep[e.g.][]{Spezzano2016}. 
These structures are believed to transport fresh, chemically young gas into the north and north-west of the core \citep{Giers2025}. 
This replenishes the abundance of \prop{} and causes the molecular peak away from the carbon-chain peak in the south.

In contrast to \prop{}, the emission of the deuterated isotopologue \propd{} exhibits two strong peaks, one located in the southern part of the core at the carbon-chain peak, similar to the emission of \dcyano{}.
The second is close to the dust peak, shifted slightly to the north-west, and coincides with the emission peak of HDCO, singly deuterated H$_2$CO \citep[see][]{ChaconTanarro2019}. 
The resulting deuterium fraction map of \propd{} peaks close to the core centre, slightly offset towards the \meth{} emission peak in the north-east with maximum values of 0.09(2). 
A similar behaviour was also observed for the carbon chain \cyc{}, where the deuteration is most efficient towards the dust peak in the centre of the core \citep{Giers2022}. 
This, together with the similarities to \dcyano{} and HDCO, supports the theory that the CCH group of \prop{} is deuterated in the gas phase, as stated in \cite{Markwick2005}.

The distribution of \dprop{} across L1544 mostly follows the emission of the main species. 
In the resulting deuteration map, however, the highest values (0.15(3)) are concentrated off-centre and towards the north-east of the core, where \meth{} peaks \citep[see e.g.][]{Spezzano2016}. 
In this region, \meth{} traces a local density enhancement \citep{Lin2022}, a clumpy substructure that introduces a deviation from the Bonnor-Ebert density profile used to describe the core structure in \cite{Keto2015}. 
In fact, the \meth{} peak seems to be the meeting point of the two larger-scale filamentary structures where L1544 is embedded. This intersection of filaments is believed to induce slow collision shocks, causing the local density enhancement \citep{Lin2022}.
The peak of the \dprop{} deuteration map overlaps with this higher-density region. 
The higher density in this area leads to a more efficient freeze-out of molecules, including CO, which drives more deuteration in the gas phase \citep[e.g.][]{Caselli2002b}. 
In addition, the enhanced freeze-out onto grains leads to higher reactivity and hence drives the reactive desorption of species from dust grains \citep[e.g.][]{Vasyunin2017}. 
The combination of these effects likely leads to the higher deuteration efficiency observed for \dprop{}.

\begin{figure}[h]
    \centering
    \includegraphics[width=\hsize]{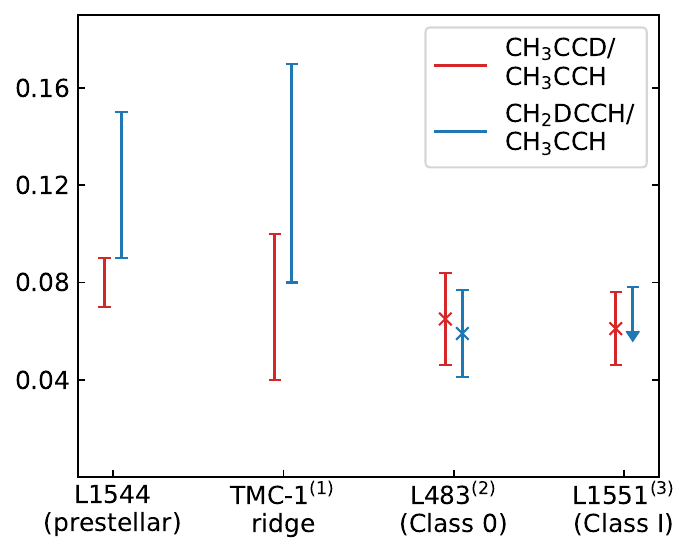}
    \caption{Comparison of the deuteration of \prop{} towards different cores. For L1544 and TMC-1, the given range covers minimum to maximum values across the cores, while for L483 and L1551 single values are given including error bars or upper limits. References include: (1) \cite{Markwick2005}, (2) \cite{Agundez2019}, (3) \cite{Marchand2024}.}
    \label{fig:ComparisonDHofCH3CCH}
\end{figure}

In general, \dprop{} shows higher levels of deuteration (0.09-0.15) than \propd{} (0.07-0.09).
Figure~\ref{fig:ComparisonDHofCH3CCH} compares the observed deuteration values of the two isotopologues to those in other cores.
Notably, the dark cloud TMC-1 shows similar behaviour to L1544: along the TMC-1 ridge, \propd{}/\prop{} shows higher values than \dprop{}/\prop{} \citep{Markwick2005}. 
In contrast, the protostellar cores L483 and L1551 show similar levels of deuteration in both isotopologues \citep{Agundez2019,Marchand2024}.

A comparison between the two deuterated isotopologues (see right-hand plot in Fig.~\ref{fig:deuterationmapsCH3CCH}) shows that in L1544 \dprop{} is 4.1-5.5 times more abundant than \propd{}.
In contrast, \cite{Markwick2005} reported a \dprop{}/\propd{} ratio ranging between 1.2–2 along the TMC-1 ridge, which is significantly lower than the statistical value of 3. 
The authors attribute these lower values to decreased abundances of H$_2$D$^+$ due to ambipolar diffusion, leading to a lower deuterium fractionation. 
On the other hand, \cite{Agundez2019} report a ratio of $3\pm1$ for $\rm CH_2DCCH/CH_3CCD$ in the dense core L483. 
The authors are able to reproduce the observed abundances of \prop{} and its singly and doubly deuterated isotopologues in this core with a pure gas-phase chemical model \citep{Agundez2021}. 
From this, they conclude that the formation and deuteration of \prop{} in L483 occurs in the gas-phase rather than on grain surfaces. 
They propose that deuterated \prop{} is formed by the dissociative recombination of C$_3$H$_5$D$^+$ and C$_3$H$_6$D$^+$.
However, in L1544 the average \dprop{}/\propd{} ratio observed is 4.5(8), which is significantly higher than the expected statistical value of 3, showing that \dprop{} has a higher deuteration efficiency than \propd{} in this core. 
Therefore, additional processes must exist that enhance deuteration of the \prop{} methyl group, or decrease the abundance of \propd{}.
Together with the tentative gradient of \dprop{}/\propd{} across the core, this suggests different deuteration mechanisms for the two functional groups with varying efficiency across the core.

A comparison of the emission of deuterated \prop{} and  deuterated methanol shows interesting similarities:
both \dprop{} and CH$_2$DOH peak in the northern part of L1544, close to the dust peak \citep[see][]{ChaconTanarro2019}, and show significantly greater deuteration in the methyl group than in the CCH or the OH group \citep[$\rm CH_2DOH/CH_3OD\geq10$,][]{Bizzocchi2014}.
The root of this might be a common formation path for \dprop{} and CH$_2$DOH on the surfaces of dust grains, which was not considered in the pure gas-phase model used in \cite{Agundez2021}.
The similarities between the \propd{} and HDCO emission provides another hint, as H$_2$CO also can be deuterated both in the gas phase and on grain surfaces.
However, extensive chemical models that incorporate both gas-phase and grain-surface reactions of large hydrocarbons, are necessary to study the chemical processes and clarify the relevance of these observed trends for the isotopic fractionation of the same functional group in different molecules.

\section{Conclusions}\label{sec:conclusion}

In this work, we studied the deuterium fraction of the carbon chains \cyano{} and \prop{} in the pre-stellar core L1544. 
We analysed the emission of the species together with their singly deuterated isotopologues and \xcyano{}. 
We derived the column densities of the different species, constraining them with non-LTE radiative transfer modelling of \cyano{} and \prop{}.
The corresponding deuterium fraction maps show different behaviours in terms of distribution and level of deuteration and seem to trace different physical conditions.

\cyano{} shows moderate levels of deuteration, with a very extended, homogeneous distribution.
This suggests that \cyano{} and \dcyano{} trace outer layers in the envelope of the core with intermediate-density gas rather than the dense core centre, and supports a gas-phase formation of \dcyano{}. 
With non-LTE modelling, we derive a $^{12}$C/$^{13}$C ratio of $60\pm20$ for \cyano{} towards the dust peak of L1544, which is consistent with the value for the local ISM, 68.
\dprop{} shows a clear deuteration peak away from the core centre, towards the north-east, coinciding with the \meth{} molecular peak. This is likely linked to the local density enhancement in this region that enhances the CO freeze-out and therefore increases deuteration and reactive desorption.
\propd{}, on the other hand, seems to be most efficiently formed closer to the centre of the core, where deuteration is also enhanced in other carbon chain such as \cyc\,.
\dprop{} shows significantly higher levels of deuteration than \propd{} and is about four to five times more abundant across the core, similar to that observed for deuterated \meth{}. 
This, together with the tentative trend of higher values towards the north, suggests different deuteration mechanisms for the two functional groups (CH$_3$ and CCH) in L1544, with varying efficiency across the core.

Overall, the results of this work suggest that gas-phase reactions dominate the formation and deuteration of carbon chains in L1544, with spatial variations driven by physical structure, density, and external radiation.
However, given the similarities of the deuterium fraction of \prop{} and \meth{}, an additional deuteration pathway of \prop{} might exist on the surfaces of dust grains.
This shows the importance of understanding the ongoing chemical processes when using deuterated molecules as tools to study the interface between the dense core and the surrounding cloud.

\begin{acknowledgements}
    We wish to thank the anonymous referee for their constructive comments.
    The authors wish to thank the Max Planck Society for financial support.
\end{acknowledgements}

%
\bibliographystyle{aa} 
\bibliography{mybib.bib} 

@ARTICLE{Agundez2019,
       author = {{Ag{\'u}ndez}, M. and {Marcelino}, N. and {Cernicharo}, J. and {Roueff}, E. and {Tafalla}, M.},
        title = "{A sensitive {\ensuremath{\lambda}} 3 mm line survey of L483. A broad view of the chemical composition of a core around a Class 0 object}",
      journal = {\aap},
         year = 2019,
        month = may,
       volume = {625},
          eid = {A147},
        pages = {A147},
          doi = {10.1051/0004-6361/201935164},
archivePrefix = {arXiv},
       eprint = {1904.06565},
 primaryClass = {astro-ph.GA},
       adsurl = {https://ui.adsabs.harvard.edu/abs/2019A&A...625A.147A}
}

@ARTICLE{Agundez2021,
       author = {{Ag{\'u}ndez}, M. and {Roueff}, E. and {Cabezas}, C. and {Cernicharo}, J. and {Marcelino}, N.},
        title = "{First detection of doubly deuterated methyl acetylene (CHD$_{2}$CCH and CH$_{2}$DCCD)}",
      journal = {\aap},
         year = 2021,
        month = may,
       volume = {649},
          eid = {A171},
        pages = {A171},
          doi = {10.1051/0004-6361/202140843},
archivePrefix = {arXiv},
       eprint = {2104.04374},
 primaryClass = {astro-ph.GA},
       adsurl = {https://ui.adsabs.harvard.edu/abs/2021A&A...649A.171A}
}

@ARTICLE{BenKhalifa2024,
       author = {{Ben Khalifa}, M. and {Darna}, B. and {Loreau}, J.},
        title = "{Collisional excitation of propyne (CH$_{3}$CCH) by He atoms}",
      journal = {\aap},
         year = 2024,
        month = mar,
       volume = {683},
          eid = {A53},
        pages = {A53},
          doi = {10.1051/0004-6361/202348717},
archivePrefix = {arXiv},
       eprint = {2402.17491},
 primaryClass = {astro-ph.GA},
       adsurl = {https://ui.adsabs.harvard.edu/abs/2024A&A...683A..53B}
}

@ARTICLE{Bianchi2023,
       author = {{Bianchi}, Eleonora and {Remijan}, Anthony and {Codella}, Claudio and {Ceccarelli}, Cecilia and {Lique}, Francois and {Spezzano}, Silvia and {Balucani}, Nadia and {Caselli}, Paola and {Herbst}, Eric and {Podio}, Linda and {Vastel}, Charlotte and {McGuire}, Brett},
        title = "{Cyanopolyyne Chemistry in the L1544 Prestellar Core: New Insights from GBT Observations}",
      journal = {\apj},
         year = 2023,
        month = feb,
       volume = {944},
       number = {2},
          eid = {208},
        pages = {208},
          doi = {10.3847/1538-4357/acb5e8},
archivePrefix = {arXiv},
       eprint = {2301.10106},
 primaryClass = {astro-ph.GA},
       adsurl = {https://ui.adsabs.harvard.edu/abs/2023ApJ...944..208B}
}

@ARTICLE{Bizzocchi2014,
       author = {{Bizzocchi}, L. and {Caselli}, P. and {Spezzano}, S. and {Leonardo}, E.},
        title = "{Deuterated methanol in the pre-stellar core L1544}",
      journal = {\aap},
         year = 2014,
        month = sep,
       volume = {569},
          eid = {A27},
        pages = {A27},
          doi = {10.1051/0004-6361/201423858},
archivePrefix = {arXiv},
       eprint = {1408.2491},
 primaryClass = {astro-ph.SR},
       adsurl = {https://ui.adsabs.harvard.edu/abs/2014A&A...569A..27B}
}

@ARTICLE{Bonnor1956,
       author = {{Bonnor}, W.~B.},
        title = "{Boyle's Law and gravitational instability}",
      journal = {\mnras},
         year = 1956,
        month = jan,
       volume = {116},
        pages = {351},
          doi = {10.1093/mnras/116.3.351},
       adsurl = {https://ui.adsabs.harvard.edu/abs/1956MNRAS.116..351B}
}

@INPROCEEDINGS{BuhlSnyder1973,
       author = {{Buhl}, David and {Snyder}, Lewis E.},
        title = "{The Detection of MM-Wave Transition of Methlacetylene}",
    booktitle = {Molecules in the Galactic Environment},
         year = 1973,
       editor = {{Gordon}, M.~A. and {Snyder}, Lewis E.},
        month = jan,
        pages = {187},
       adsurl = {https://ui.adsabs.harvard.edu/abs/1973mge..conf..187B}
}

@ARTICLE{Calcutt2019,
       author = {{Calcutt}, H. and {Willis}, E.~R. and {J{\o}rgensen}, J.~K. and {Bjerkeli}, P. and {Ligterink}, N.~F.~W. and {Coutens}, A. and {M{\"u}ller}, H.~S.~P. and {Garrod}, R.~T. and {Wampfler}, S.~F. and {Drozdovskaya}, M.~N.},
        title = "{The ALMA-PILS survey: propyne (CH$_{3}$CCH) in IRAS 16293-2422}",
      journal = {\aap},
         year = 2019,
        month = nov,
       volume = {631},
          eid = {A137},
        pages = {A137},
          doi = {10.1051/0004-6361/201936323},
archivePrefix = {arXiv},
       eprint = {1909.13329},
 primaryClass = {astro-ph.SR},
       adsurl = {https://ui.adsabs.harvard.edu/abs/2019A&A...631A.137C}
}

@ARTICLE{Caselli1999,
       author = {{Caselli}, P. and {Walmsley}, C.~M. and {Tafalla}, M. and {Dore}, L. and {Myers}, P.~C.},
        title = "{CO Depletion in the Starless Cloud Core L1544}",
      journal = {\apjl},
         year = 1999,
        month = oct,
       volume = {523},
       number = {2},
        pages = {L165-L169},
          doi = {10.1086/312280},
       adsurl = {https://ui.adsabs.harvard.edu/abs/1999ApJ...523L.165C}
}

@ARTICLE{Caselli2002a,
       author = {{Caselli}, P. and {Walmsley}, C.~M. and {Zucconi}, A. and {Tafalla}, M. and {Dore}, L. and {Myers}, P.~C.},
        title = "{Molecular Ions in L1544. I. Kinematics}",
      journal = {\apj},
         year = 2002,
        month = jan,
       volume = {565},
       number = {1},
        pages = {331-343},
          doi = {10.1086/324301},
archivePrefix = {arXiv},
       eprint = {astro-ph/0109021},
 primaryClass = {astro-ph},
       adsurl = {https://ui.adsabs.harvard.edu/abs/2002ApJ...565..331C}
}

@ARTICLE{Caselli2002b,
       author = {{Caselli}, P. and {Walmsley}, C.~M. and {Zucconi}, A. and {Tafalla}, M. and {Dore}, L. and {Myers}, P.~C.},
        title = "{Molecular Ions in L1544. II. The Ionization Degree}",
      journal = {\apj},
         year = 2002,
        month = jan,
       volume = {565},
       number = {1},
        pages = {344-358},
          doi = {10.1086/324302},
archivePrefix = {arXiv},
       eprint = {astro-ph/0109023},
 primaryClass = {astro-ph},
       adsurl = {https://ui.adsabs.harvard.edu/abs/2002ApJ...565..344C}
}

@ARTICLE{Caselli2022,
       author = {{Caselli}, Paola and {Pineda}, Jaime E. and {Sipil{\"a}}, Olli and {Zhao}, Bo and {Redaelli}, Elena and {Spezzano}, Silvia and {Maureira}, Maria Jos{\'e} and {Alves}, Felipe and {Bizzocchi}, Luca and {Bourke}, Tyler L. and {Chac{\'o}n-Tanarro}, Ana and {Friesen}, Rachel and {Galli}, Daniele and {Harju}, Jorma and {Jim{\'e}nez-Serra}, Izaskun and {Keto}, Eric and {Li}, Zhi-Yun and {Padovani}, Marco and {Schmiedeke}, Anika and {Tafalla}, Mario and {Vastel}, Charlotte},
        title = "{The Central 1000 au of a Prestellar Core Revealed with ALMA. II. Almost Complete Freeze-out}",
      journal = {\apj},
         year = 2022,
        month = apr,
       volume = {929},
       number = {1},
          eid = {13},
        pages = {13},
          doi = {10.3847/1538-4357/ac5913},
archivePrefix = {arXiv},
       eprint = {2202.13374},
 primaryClass = {astro-ph.SR},
       adsurl = {https://ui.adsabs.harvard.edu/abs/2022ApJ...929...13C}
}

@ARTICLE{ChaconTanarro2019,
       author = {{Chac{\'o}n-Tanarro}, A. and {Caselli}, P. and {Bizzocchi}, L. and {Pineda}, J.~E. and {Sipil{\"a}}, O. and {Vasyunin}, A. and {Spezzano}, S. and {Punanova}, A. and {Giuliano}, B.~M. and {Lattanzi}, V.},
        title = "{Mapping deuterated methanol toward L1544. I. Deuterium fraction and comparison with modeling}",
      journal = {\aap},
         year = 2019,
        month = feb,
       volume = {622},
          eid = {A141},
        pages = {A141},
          doi = {10.1051/0004-6361/201832703},
archivePrefix = {arXiv},
       eprint = {1808.09871},
 primaryClass = {astro-ph.GA},
       adsurl = {https://ui.adsabs.harvard.edu/abs/2019A&A...622A.141C}
}

@ARTICLE{Chantzos2018,
       author = {{Chantzos}, J. and {Spezzano}, S. and {Caselli}, P. and {Chac{\'o}n-Tanarro}, A. and {Bizzocchi}, L. and {Sipil{\"a}}, O. and {Giuliano}, B.~M.},
        title = "{A Study of the c-C$_{3}$HD/c-C$_{3}$H$_{2}$ Ratio in Low-mass Star-forming Regions}",
      journal = {\apj},
         year = 2018,
        month = aug,
       volume = {863},
       number = {2},
          eid = {126},
        pages = {126},
          doi = {10.3847/1538-4357/aad2dc},
archivePrefix = {arXiv},
       eprint = {1807.04663},
 primaryClass = {astro-ph.SR},
       adsurl = {https://ui.adsabs.harvard.edu/abs/2018ApJ...863..126C}
}

@ARTICLE{Chapillon2012,
       author = {{Chapillon}, Edwige and {Dutrey}, Anne and {Guilloteau}, St{\'e}phane and {Pi{\'e}tu}, Vincent and {Wakelam}, Valentine and {Hersant}, Franck and {Gueth}, Fr{\'e}deric and {Henning}, Thomas and {Launhardt}, Ralf and {Schreyer}, Katharina and {Semenov}, Dmitry},
        title = "{Chemistry in Disks. VII. First Detection of HC$_{3}$N in Protoplanetary Disks}",
      journal = {\apj},
         year = 2012,
        month = sep,
       volume = {756},
       number = {1},
          eid = {58},
        pages = {58},
          doi = {10.1088/0004-637X/756/1/58},
archivePrefix = {arXiv},
       eprint = {1207.2682},
 primaryClass = {astro-ph.SR},
       adsurl = {https://ui.adsabs.harvard.edu/abs/2012ApJ...756...58C}
}

@ARTICLE{Crapsi2005,
       author = {{Crapsi}, A. and {Caselli}, P. and {Walmsley}, C.~M. and {Myers}, P.~C. and {Tafalla}, M. and {Lee}, C.~W. and {Bourke}, T.~L.},
        title = "{Probing the Evolutionary Status of Starless Cores through N$_{2}$H$^{+}$ and N$_{2}$D$^{+}$ Observations}",
      journal = {\apj},
         year = 2005,
        month = jan,
       volume = {619},
       number = {1},
        pages = {379-406},
          doi = {10.1086/426472},
archivePrefix = {arXiv},
       eprint = {astro-ph/0409529},
 primaryClass = {astro-ph},
       adsurl = {https://ui.adsabs.harvard.edu/abs/2005ApJ...619..379C}
}

@ARTICLE{DalgarnoLepp1984,
       author = {{Dalgarno}, A. and {Lepp}, S.},
        title = "{Deuterium fractionation mechanisms in interstellar clouds.}",
      journal = {\apjl},
         year = 1984,
        month = dec,
       volume = {287},
        pages = {L47-L50},
          doi = {10.1086/184395},
       adsurl = {https://ui.adsabs.harvard.edu/abs/1984ApJ...287L..47D}
}

@ARTICLE{Ebert1955,
       author = {{Ebert}, R.},
        title = "{{\"U}ber die Verdichtung von H I-Gebieten. Mit 5 Textabbildungen}",
      journal = {\zap},
         year = 1955,
        month = jan,
       volume = {37},
        pages = {217},
       adsurl = {https://ui.adsabs.harvard.edu/abs/1955ZA.....37..217E}
}

@ARTICLE{Faure2016,
       author = {{Faure}, Alexandre and {Lique}, Fran{\c{c}}ois and {Wiesenfeld}, Laurent},
        title = "{Collisional excitation of HC$_{3}$N by para- and ortho-H$_{2}$}",
      journal = {\mnras},
         year = 2016,
        month = aug,
       volume = {460},
       number = {2},
        pages = {2103-2109},
          doi = {10.1093/mnras/stw1156},
archivePrefix = {arXiv},
       eprint = {1605.03786},
 primaryClass = {astro-ph.GA},
       adsurl = {https://ui.adsabs.harvard.edu/abs/2016MNRAS.460.2103F}
}

@ARTICLE{Fontani2002,
       author = {{Fontani}, F. and {Cesaroni}, R. and {Caselli}, P. and {Olmi}, L.},
        title = "{The structure of molecular clumps around high-mass young stellar objects}",
      journal = {\aap},
         year = 2002,
        month = jul,
       volume = {389},
        pages = {603-617},
          doi = {10.1051/0004-6361:20020579},
archivePrefix = {arXiv},
       eprint = {astro-ph/0204313},
 primaryClass = {astro-ph},
       adsurl = {https://ui.adsabs.harvard.edu/abs/2002A&A...389..603F}
}

@ARTICLE{FullerMyers1992,
       author = {{Fuller}, G.~A. and {Myers}, P.~C.},
        title = "{Dense Cores in Dark Clouds. VII. Line Width--Size Relations}",
      journal = {\apj},
         year = 1992,
        month = jan,
       volume = {384},
        pages = {523},
          doi = {10.1086/170894},
       adsurl = {https://ui.adsabs.harvard.edu/abs/1992ApJ...384..523F}
}

@ARTICLE{Gerin1992,
       author = {{Gerin}, M. and {Combes}, F. and {Wlodarczak}, G. and {Encrenaz}, P. and {Laurent}, C.},
        title = "{Interstellar detection of deuterated methyl acetylene.}",
      journal = {\aap},
         year = 1992,
        month = jan,
       volume = {253},
        pages = {L29-L32},
       adsurl = {https://ui.adsabs.harvard.edu/abs/1992A&A...253L..29G}
}

@ARTICLE{Giers2022,
       author = {{Giers}, K. and {Spezzano}, S. and {Alves}, F. and {Caselli}, P. and {Redaelli}, E. and {Sipil{\"a}}, O. and {Ben Khalifa}, M. and {Wiesenfeld}, L. and {Br{\"u}nken}, S. and {Bizzocchi}, L.},
        title = "{Deuteration of c-C$_{3}$H$_{2}$ towards the pre-stellar core L1544}",
      journal = {\aap},
         year = 2022,
        month = aug,
       volume = {664},
          eid = {A119},
        pages = {A119},
          doi = {10.1051/0004-6361/202243422},
archivePrefix = {arXiv},
       eprint = {2205.14963},
 primaryClass = {astro-ph.GA},
       adsurl = {https://ui.adsabs.harvard.edu/abs/2022A&A...664A.119G}
}

@ARTICLE{Giers2023,
       author = {{Giers}, K. and {Spezzano}, S. and {Caselli}, P. and {Wirstr{\"o}m}, E. and {Sipil{\"a}}, O. and {Pineda}, J.~E. and {Redaelli}, E. and {Bop}, C.~T. and {Lique}, F.},
        title = "{Similar levels of deuteration in the pre-stellar core L1544 and the protostellar core HH211}",
      journal = {\aap},
         year = 2023,
        month = aug,
       volume = {676},
          eid = {A78},
        pages = {A78},
          doi = {10.1051/0004-6361/202346433},
archivePrefix = {arXiv},
       eprint = {2306.12775},
 primaryClass = {astro-ph.GA},
       adsurl = {https://ui.adsabs.harvard.edu/abs/2023A&A...676A..78G}
}

@ARTICLE{Giers2025,
       author = {{Giers}, K. and {Spezzano}, S. and {Lin}, Y. and {Valdivia-Mena}, M.~T. and {Caselli}, P. and {Sipil{\"a}}, O.},
        title = "{Chemical segregation analysed with unsupervised clustering}",
      journal = {\aap},
         year = 2025,
        month = jul,
       volume = {699},
          eid = {A103},
        pages = {A103},
          doi = {10.1051/0004-6361/202553843},
       adsurl = {https://ui.adsabs.harvard.edu/abs/2025A&A...699A.103G}
}

@ARTICLE{GoldsmithLanger1999,
       author = {{Goldsmith}, Paul F. and {Langer}, William D.},
        title = "{Population Diagram Analysis of Molecular Line Emission}",
      journal = {\apj},
         year = 1999,
        month = may,
       volume = {517},
       number = {1},
        pages = {209-225},
          doi = {10.1086/307195},
       adsurl = {https://ui.adsabs.harvard.edu/abs/1999ApJ...517..209G}
}

@ARTICLE{Gratier2016,
       author = {{Gratier}, P. and {Majumdar}, L. and {Ohishi}, M. and {Roueff}, E. and {Loison}, J.~C. and {Hickson}, K.~M. and {Wakelam}, V.},
        title = "{A New Reference Chemical Composition for TMC-1}",
      journal = {\apjs},
         year = 2016,
        month = aug,
       volume = {225},
       number = {2},
          eid = {25},
        pages = {25},
          doi = {10.3847/0067-0049/225/2/25},
archivePrefix = {arXiv},
       eprint = {1610.00524},
 primaryClass = {astro-ph.GA},
       adsurl = {https://ui.adsabs.harvard.edu/abs/2016ApJS..225...25G}
}

@ARTICLE{Guzman2018,
       author = {{Guzm{\'a}n}, Andr{\'e}s E. and {Guzm{\'a}n}, Viviana V. and {Garay}, Guido and {Bronfman}, Leonardo and {Hechenleitner}, Federico},
        title = "{Chemistry of the High-mass Protostellar Molecular Clump IRAS 16562-3959}",
      journal = {\apjs},
         year = 2018,
        month = jun,
       volume = {236},
       number = {2},
          eid = {45},
        pages = {45},
          doi = {10.3847/1538-4365/aac01d},
archivePrefix = {arXiv},
       eprint = {1804.06544},
 primaryClass = {astro-ph.GA},
       adsurl = {https://ui.adsabs.harvard.edu/abs/2018ApJS..236...45G}
}

@ARTICLE{Hickson2016,
       author = {{Hickson}, Kevin M. and {Wakelam}, Valentine and {Loison}, Jean-Christophe},
        title = "{Methylacetylene (CH$_{3}$CCH) and propene (C$_{3}$H$_{6}$) formation in cold dense clouds: A case of dust grain chemistry}",
      journal = {Molecular Astrophysics},
         year = 2016,
        month = apr,
       volume = {3},
        pages = {1-9},
          doi = {10.1016/j.molap.2016.03.001},
archivePrefix = {arXiv},
       eprint = {1603.02703},
 primaryClass = {astro-ph.GA},
       adsurl = {https://ui.adsabs.harvard.edu/abs/2016MolAs...3....1H}
}

@ARTICLE{HilyBlant2018b,
       author = {{Hily-Blant}, P. and {Faure}, A. and {Vastel}, C. and {Magalhaes}, V. and {Lefloch}, B. and {Bachiller}, R.},
        title = "{The nitrogen isotopic ratio of HC$_{3}$N towards the L1544 prestellar core}",
      journal = {\mnras},
         year = 2018,
        month = oct,
       volume = {480},
       number = {1},
        pages = {1174-1186},
          doi = {10.1093/mnras/sty1562},
archivePrefix = {arXiv},
       eprint = {1806.04428},
 primaryClass = {astro-ph.GA},
       adsurl = {https://ui.adsabs.harvard.edu/abs/2018MNRAS.480.1174H}
}

@ARTICLE{Howe1994,
       author = {{Howe}, D.~A. and {Millar}, T.~J. and {Schilke}, P. and {Walmsley}, C.~M.},
        title = "{Observations of deuterated cyanoacetylene in dark clouds.}",
      journal = {\mnras},
         year = 1994,
        month = mar,
       volume = {267},
        pages = {59-68},
          doi = {10.1093/mnras/267.1.59},
       adsurl = {https://ui.adsabs.harvard.edu/abs/1994MNRAS.267...59H}
}

@ARTICLE{Juvela2020,
       author = {{Juvela}, Mika},
        title = "{LOC program for line radiative transfer}",
      journal = {\aap},
         year = 2020,
        month = dec,
       volume = {644},
          eid = {A151},
        pages = {A151},
          doi = {10.1051/0004-6361/202039456},
archivePrefix = {arXiv},
       eprint = {2009.12609},
 primaryClass = {astro-ph.IM},
       adsurl = {https://ui.adsabs.harvard.edu/abs/2020A&A...644A.151J}
}

@ARTICLE{Keto2015,
       author = {{Keto}, Eric and {Caselli}, Paola and {Rawlings}, Jonathan},
        title = "{The dynamics of collapsing cores and star formation}",
      journal = {\mnras},
         year = 2015,
        month = feb,
       volume = {446},
       number = {4},
        pages = {3731-3740},
          doi = {10.1093/mnras/stu2247},
archivePrefix = {arXiv},
       eprint = {1410.5889},
 primaryClass = {astro-ph.SR},
       adsurl = {https://ui.adsabs.harvard.edu/abs/2015MNRAS.446.3731K}
}

@ARTICLE{Langer1980,
       author = {{Langer}, W.~D. and {Schloerb}, F.~P. and {Snell}, R.~L. and {Young}, J.~S.},
        title = "{Detection of deuterated cyanoacetylene in the interstellar cloud TMC 1}",
      journal = {\apjl},
         year = 1980,
        month = aug,
       volume = {239},
        pages = {L125-L128},
          doi = {10.1086/183307},
       adsurl = {https://ui.adsabs.harvard.edu/abs/1980ApJ...239L.125L}
}

@ARTICLE{LinWyrowski2022,
       author = {{Lin}, Y. and {Wyrowski}, F. and {Liu}, H.~B. and {Izquierdo}, A.~F. and {Csengeri}, T. and {Leurini}, S. and {Menten}, K.~M.},
        title = "{The evolution of temperature and density structures of OB cluster-forming molecular clumps}",
      journal = {\aap},
         year = 2022,
        month = feb,
       volume = {658},
          eid = {A128},
        pages = {A128},
          doi = {10.1051/0004-6361/202142023},
archivePrefix = {arXiv},
       eprint = {2112.01115},
 primaryClass = {astro-ph.GA},
       adsurl = {https://ui.adsabs.harvard.edu/abs/2022A&A...658A.128L}
}

@ARTICLE{Lin2022,
       author = {{Lin}, Y. and {Spezzano}, S. and {Sipil{\"a}}, O. and {Vasyunin}, A. and {Caselli}, P.},
        title = "{Multiline observations of CH$_{3}$OH, c-C$_{3}$H$_{2}$, and HNCO toward L1544. Dissecting the core structure with chemical differentiation}",
      journal = {\aap},
         year = 2022,
        month = sep,
       volume = {665},
          eid = {A131},
        pages = {A131},
          doi = {10.1051/0004-6361/202243657},
archivePrefix = {arXiv},
       eprint = {2205.09806},
 primaryClass = {astro-ph.GA},
       adsurl = {https://ui.adsabs.harvard.edu/abs/2022A&A...665A.131L}
}

@ARTICLE{Mangum2015,
       author = {{Mangum}, Jeffrey G. and {Shirley}, Yancy L.},
        title = "{How to Calculate Molecular Column Density}",
      journal = {\pasp},
         year = 2015,
        month = mar,
       volume = {127},
       number = {949},
        pages = {266},
          doi = {10.1086/680323},
archivePrefix = {arXiv},
       eprint = {1501.01703},
 primaryClass = {astro-ph.IM},
       adsurl = {https://ui.adsabs.harvard.edu/abs/2015PASP..127..266M}
}

@ARTICLE{Marchand2024,
       author = {{Marchand}, P. and {Coutens}, A. and {Scigliuto}, J. and {de Miera}, F. Cruz-S{\'a}enz and {Andreu}, A. and {Loison}, J. -C. and {K{\'o}sp{\'a}l}, {\'A}. and {{\'A}br{\'a}ham}, P.},
        title = "{Chemical inventory of the envelope of the Class I protostar L1551 IRS 5}",
      journal = {\aap},
         year = 2024,
        month = jul,
       volume = {687},
          eid = {A195},
        pages = {A195},
          doi = {10.1051/0004-6361/202450023},
archivePrefix = {arXiv},
       eprint = {2405.08517},
 primaryClass = {astro-ph.GA},
       adsurl = {https://ui.adsabs.harvard.edu/abs/2024A&A...687A.195M}
}

@ARTICLE{Markwick2005,
       author = {{Markwick}, A.~J. and {Charnley}, S.~B. and {Butner}, H.~M. and {Millar}, T.~J.},
        title = "{Interstellar CH$_{3}$CCD}",
      journal = {\apjl},
         year = 2005,
        month = jul,
       volume = {627},
       number = {2},
        pages = {L117-L120},
          doi = {10.1086/432415},
       adsurl = {https://ui.adsabs.harvard.edu/abs/2005ApJ...627L.117M}
}

@ARTICLE{Milam2005,
       author = {{Milam}, S.~N. and {Savage}, C. and {Brewster}, M.~A. and {Ziurys}, L.~M. and {Wyckoff}, S.},
        title = "{The $^{12}$C/$^{13}$C Isotope Gradient Derived from Millimeter Transitions of CN: The Case for Galactic Chemical Evolution}",
      journal = {\apj},
         year = 2005,
        month = dec,
       volume = {634},
       number = {2},
        pages = {1126-1132},
          doi = {10.1086/497123},
       adsurl = {https://ui.adsabs.harvard.edu/abs/2005ApJ...634.1126M}
}

@ARTICLE{Morris1976,
       author = {{Morris}, M. and {Turner}, B.~E. and {Palmer}, P. and {Zuckerman}, B.},
        title = "{Cyanoacetylene in dense interstellar clouds.}",
      journal = {\apj},
         year = 1976,
        month = apr,
       volume = {205},
        pages = {82-93},
          doi = {10.1086/154252},
       adsurl = {https://ui.adsabs.harvard.edu/abs/1976ApJ...205...82M}
}

@ARTICLE{Mueller2001,
       author = {{M{\"u}ller}, H.~S.~P. and {Thorwirth}, S. and {Roth}, D.~A. and {Winnewisser}, G.},
        title = "{The Cologne Database for Molecular Spectroscopy, CDMS}",
      journal = {\aap},
         year = 2001,
        month = apr,
       volume = {370},
        pages = {L49-L52},
          doi = {10.1051/0004-6361:20010367},
       adsurl = {https://ui.adsabs.harvard.edu/abs/2001A&A...370L..49M}
}

@ARTICLE{Oberg2013,
       author = {{{\"O}berg}, Karin I. and {Boamah}, Mavis D. and {Fayolle}, Edith C. and {Garrod}, Robin T. and {Cyganowski}, Claudia J. and {van der Tak}, Floris},
        title = "{The Spatial Distribution of Organics toward the High-mass YSO NGC 7538 IRS9}",
      journal = {\apj},
         year = 2013,
        month = jul,
       volume = {771},
       number = {2},
          eid = {95},
        pages = {95},
          doi = {10.1088/0004-637X/771/2/95},
archivePrefix = {arXiv},
       eprint = {1305.3151},
 primaryClass = {astro-ph.GA},
       adsurl = {https://ui.adsabs.harvard.edu/abs/2013ApJ...771...95O}
}

@ARTICLE{Pagani1992,
       author = {{Pagani}, L. and {Salez}, M. and {Wannier}, P.~G.},
        title = "{The chemistry of H2D+ in cold clouds.}",
      journal = {\aap},
         year = 1992,
        month = may,
       volume = {258},
        pages = {479-488},
       adsurl = {https://ui.adsabs.harvard.edu/abs/1992A&A...258..479P}
}

@InProceedings{pandas,
  author    = { {W}es {M}c{K}inney },
  title     = { {D}ata {S}tructures for {S}tatistical {C}omputing in {P}ython },
  booktitle = { {P}roceedings of the 9th {P}ython in {S}cience {C}onference },
  pages     = { 56 - 61 },
  year      = { 2010 },
  editor    = { {S}t\'efan van der {W}alt and {J}arrod {M}illman },
  doi       = { 10.25080/Majora-92bf1922-00a }
}

@Article{ParkerKaiser2017,
author ="Parker, Dorian S. N. and Kaiser, Ralf I.",
title  ="On the formation of nitrogen-substituted polycyclic aromatic hydrocarbons (NPAHs) in circumstellar and interstellar environments",
journal  ="Chem. Soc. Rev.",
year  ="2017",
volume  ="46",
issue  ="2",
pages  ="452-463",
publisher  ="The Royal Society of Chemistry",
doi  ="10.1039/C6CS00714G",
url  ="http://dx.doi.org/10.1039/C6CS00714G"}

@INPROCEEDINGS{Pety2005,
       author = {{Pety}, J.},
        title = "{Successes of and Challenges to GILDAS, a State-of-the-Art Radioastronomy Toolkit}",
    booktitle = {SF2A-2005: Semaine de l'Astrophysique Francaise},
         year = 2005,
       editor = {{Casoli}, F. and {Contini}, T. and {Hameury}, J.~M. and {Pagani}, L.},
        month = dec,
        pages = {721},
       adsurl = {https://ui.adsabs.harvard.edu/abs/2005sf2a.conf..721P}
}

@MISC{pyspeckit,
       author = {{Ginsburg}, Adam and {Mirocha}, Jordan},
        title = "{PySpecKit: Python Spectroscopic Toolkit}",
 howpublished = {Astrophysics Source Code Library, record ascl:1109.001},
         year = 2011,
        month = sep,
          eid = {ascl:1109.001},
        pages = {ascl:1109.001},
archivePrefix = {ascl},
       eprint = {1109.001},
       adsurl = {https://ui.adsabs.harvard.edu/abs/2011ascl.soft09001G}
}

@ARTICLE{Redaelli2019,
       author = {{Redaelli}, E. and {Bizzocchi}, L. and {Caselli}, P. and {Sipil{\"a}}, O. and {Lattanzi}, V. and {Giuliano}, B.~M. and {Spezzano}, S.},
        title = "{High-sensitivity maps of molecular ions in L1544. I. Deuteration of N$_{2}$H$^{+}$ and HCO$^{+}$ and primary evidence of N$_{2}$D$^{+}$ depletion}",
      journal = {\aap},
         year = 2019,
        month = sep,
       volume = {629},
          eid = {A15},
        pages = {A15},
          doi = {10.1051/0004-6361/201935314},
archivePrefix = {arXiv},
       eprint = {1907.08217},
 primaryClass = {astro-ph.SR},
       adsurl = {https://ui.adsabs.harvard.edu/abs/2019A&A...629A..15R}
}

@ARTICLE{Rivilla2020,
       author = {{Rivilla}, V.~M. and {Colzi}, L. and {Fontani}, F. and {Melosso}, M. and {Caselli}, P. and {Bizzocchi}, L. and {Tamassia}, F. and {Dore}, L.},
        title = "{DC$_{3}$N observations towards high-mass star-forming regions}",
      journal = {\mnras},
         year = 2020,
        month = aug,
       volume = {496},
       number = {2},
        pages = {1990-1999},
          doi = {10.1093/mnras/staa1616},
archivePrefix = {arXiv},
       eprint = {2005.14118},
 primaryClass = {astro-ph.GA},
       adsurl = {https://ui.adsabs.harvard.edu/abs/2020MNRAS.496.1990R}
}

@ARTICLE{Sakai2009,
       author = {{Sakai}, Nami and {Sakai}, Takeshi and {Hirota}, Tomoya and {Yamamoto}, Satoshi},
        title = "{Deuterated Molecules in Warm Carbon Chain Chemistry: The L1527 Case}",
      journal = {\apj},
         year = 2009,
        month = sep,
       volume = {702},
       number = {2},
        pages = {1025-1035},
          doi = {10.1088/0004-637X/702/2/1025},
       adsurl = {https://ui.adsabs.harvard.edu/abs/2009ApJ...702.1025S}
}

@ARTICLE{SchiffBohme1979,
       author = {{Schiff}, H.~I. and {Bohme}, D.~K.},
        title = "{An ion-molecule scheme for the synthesis of hydrocarbon-chain and organonitrogen molecules in dense interstellar clouds.}",
      journal = {\apj},
         year = 1979,
        month = sep,
       volume = {232},
        pages = {740-746},
          doi = {10.1086/157334},
       adsurl = {https://ui.adsabs.harvard.edu/abs/1979ApJ...232..740S}
}

@ARTICLE{Schoier2005,
       author = {{Sch{\"o}ier}, F.~L. and {van der Tak}, F.~F.~S. and {van Dishoeck}, E.~F. and {Black}, J.~H.},
        title = "{An atomic and molecular database for analysis of submillimetre line observations}",
      journal = {\aap},
         year = 2005,
        month = mar,
       volume = {432},
       number = {1},
        pages = {369-379},
          doi = {10.1051/0004-6361:20041729},
archivePrefix = {arXiv},
       eprint = {astro-ph/0411110},
 primaryClass = {astro-ph},
       adsurl = {https://ui.adsabs.harvard.edu/abs/2005A&A...432..369S}
}

@ARTICLE{Shingledecker2021,
       author = {{Shingledecker}, C.~N. and {Lee}, K.~L.~K. and {Wandishin}, J.~T. and {Balucani}, N. and {Burkhardt}, A.~M. and {Charnley}, S.~B. and {Loomis}, R. and {Schreffler}, M. and {Siebert}, M. and {McCarthy}, M.~C. and {McGuire}, B.~A.},
        title = "{Detection of interstellar H$_{2}$CCCHC$_{3}$N. A possible link between chains and rings in cold cores}",
      journal = {\aap},
         year = 2021,
        month = aug,
       volume = {652},
          eid = {L12},
        pages = {L12},
          doi = {10.1051/0004-6361/202140698},
       adsurl = {https://ui.adsabs.harvard.edu/abs/2021A&A...652L..12S}
}

@ARTICLE{Sipila2019,
       author = {{Sipil{\"a}}, O. and {Caselli}, P. and {Redaelli}, E. and {Juvela}, M. and {Bizzocchi}, L.},
        title = "{Why does ammonia not freeze out in the centre of pre-stellar cores?}",
      journal = {\mnras},
         year = 2019,
        month = jul,
       volume = {487},
       number = {1},
        pages = {1269-1282},
          doi = {10.1093/mnras/stz1344},
archivePrefix = {arXiv},
       eprint = {1905.02384},
 primaryClass = {astro-ph.GA},
       adsurl = {https://ui.adsabs.harvard.edu/abs/2019MNRAS.487.1269S}
}

@software{Spectralcube,
       author = {{Ginsburg}, Adam and {Koch}, Eric and {Robitaille}, Thomas and {Beaumont}, Chris and {Adamginsburg} and {ZuHone}, John and {Sipocz}, Brigitta and {Patra}, Sushobhana and {Jones}, Craig and {Lim}, P.~L. and {Rosolowsky}, Erik and {Stern}, Kris and {Earl}, Nicholas and {De Val-Borro}, Miguel and {Jrobbfed} and {Shuokong} and {Kepley}, Amanda and {Sokolov}, Vlas and {Badger}, The Gitter and {Maret}, S{\'e}bastien and {Garrido}, Juli{\'a}n and {Booker}, Joseph and {Tollerud}, Erik},
        title = "{radio-astro-tools/spectral-cube: v0.4.4}",
         year = 2019,
        month = feb,
          eid = {10.5281/zenodo.2573901},
          doi = {10.5281/zenodo.2573901},
      version = {v0.4.4},
    publisher = {Zenodo},
       adsurl = {https://ui.adsabs.harvard.edu/abs/2019zndo...2573901G}
}

@INPROCEEDINGS{Spezzano2013,
       author = {{Spezzano}, Silvia and {Brunken}, Sandra and {Schilke}, Peter and {Menten}, Karl M. and {Caselli}, Paola and {McCarthy}, Michael C. and {Bizzocchi}, Luca and {Trevino}, Sandra and {Aikawa}, Yuri and {Schlemmer}, Stephan},
        title = "{Detection and Formation of Interstellar c-C\_3D\_2}",
    booktitle = {68th International Symposium on Molecular Spectroscopy},
         year = 2013,
        month = jun,
          eid = {ETI08},
        pages = {ETI08},
       adsurl = {https://ui.adsabs.harvard.edu/abs/2013mss..confETI08S}
}

@ARTICLE{Spezzano2016,
       author = {{Spezzano}, S. and {Bizzocchi}, L. and {Caselli}, P. and {Harju}, J. and {Br{\"u}nken}, S.},
        title = "{Chemical differentiation in a prestellar core traces non-uniform illumination}",
      journal = {\aap},
         year = 2016,
        month = aug,
       volume = {592},
          eid = {L11},
        pages = {L11},
          doi = {10.1051/0004-6361/201628652},
archivePrefix = {arXiv},
       eprint = {1607.03242},
 primaryClass = {astro-ph.GA},
       adsurl = {https://ui.adsabs.harvard.edu/abs/2016A&A...592L..11S}
}

@ARTICLE{Spezzano2017,
       author = {{Spezzano}, S. and {Caselli}, P. and {Bizzocchi}, L. and {Giuliano}, B.~M. and {Lattanzi}, V.},
        title = "{The observed chemical structure of L1544}",
      journal = {\aap},
         year = 2017,
        month = oct,
       volume = {606},
          eid = {A82},
        pages = {A82},
          doi = {10.1051/0004-6361/201731262},
archivePrefix = {arXiv},
       eprint = {1707.06015},
 primaryClass = {astro-ph.GA},
       adsurl = {https://ui.adsabs.harvard.edu/abs/2017A&A...606A..82S}
}

@ARTICLE{Spezzano2022,
       author = {{Spezzano}, S. and {Caselli}, P. and {Sipil{\"a}}, O. and {Bizzocchi}, L.},
        title = "{Nitrogen fractionation towards a pre-stellar core traces isotope-selective photodissociation}",
      journal = {\aap},
         year = 2022,
        month = aug,
       volume = {664},
          eid = {L2},
        pages = {L2},
          doi = {10.1051/0004-6361/202244301},
archivePrefix = {arXiv},
       eprint = {2207.06121},
 primaryClass = {astro-ph.GA},
       adsurl = {https://ui.adsabs.harvard.edu/abs/2022A&A...664L...2S}
}

@ARTICLE{Turner1971,
       author = {{Turner}, B.~E.},
        title = "{Detection of Interstellar Cyanoacetylene}",
      journal = {\apjl},
         year = 1971,
        month = jan,
       volume = {163},
        pages = {L35},
          doi = {10.1086/180662},
       adsurl = {https://ui.adsabs.harvard.edu/abs/1971ApJ...163L..35T}
}

@ARTICLE{Turner1999,
       author = {{Turner}, B.~E. and {Terzieva}, R. and {Herbst}, Eric},
        title = "{The Physics and Chemistry of Small Translucent Molecular Clouds. XII. More Complex Species Explainable by Gas-Phase Processes}",
      journal = {\apj},
         year = 1999,
        month = jun,
       volume = {518},
       number = {2},
        pages = {699-732},
          doi = {10.1086/307300},
       adsurl = {https://ui.adsabs.harvard.edu/abs/1999ApJ...518..699T}
}

@ARTICLE{Turner2001,
       author = {{Turner}, B.~E.},
        title = "{Deuterated Molecules in Translucent and Dark Clouds}",
      journal = {\apjs},
         year = 2001,
        month = oct,
       volume = {136},
       number = {2},
        pages = {579-629},
          doi = {10.1086/322536},
       adsurl = {https://ui.adsabs.harvard.edu/abs/2001ApJS..136..579T}
}

@ARTICLE{vanDishoeck1995,
       author = {{van Dishoeck}, Ewine F. and {Blake}, Geoffrey A. and {Jansen}, David J. and {Groesbeck}, T.~D.},
        title = "{Molecular Abundances and Low-Mass Star Formation. II. Organic and Deuterated Species toward IRAS 16293-2422}",
      journal = {\apj},
         year = 1995,
        month = jul,
       volume = {447},
        pages = {760},
          doi = {10.1086/175915},
       adsurl = {https://ui.adsabs.harvard.edu/abs/1995ApJ...447..760V}
}

@ARTICLE{Vastel2014,
       author = {{Vastel}, C. and {Ceccarelli}, C. and {Lefloch}, B. and {Bachiller}, R.},
        title = "{The Origin of Complex Organic Molecules in Prestellar Cores}",
      journal = {\apjl},
         year = 2014,
        month = nov,
       volume = {795},
       number = {1},
          eid = {L2},
        pages = {L2},
          doi = {10.1088/2041-8205/795/1/L2},
archivePrefix = {arXiv},
       eprint = {1409.6565},
 primaryClass = {astro-ph.SR},
       adsurl = {https://ui.adsabs.harvard.edu/abs/2014ApJ...795L...2V}
}

@ARTICLE{Vasyunin2017,
       author = {{Vasyunin}, A.~I. and {Caselli}, P. and {Dulieu}, F. and {Jim{\'e}nez-Serra}, I.},
        title = "{Formation of Complex Molecules in Prestellar Cores: A Multilayer Approach}",
      journal = {\apj},
         year = 2017,
        month = jun,
       volume = {842},
       number = {1},
          eid = {33},
        pages = {33},
          doi = {10.3847/1538-4357/aa72ec},
archivePrefix = {arXiv},
       eprint = {1705.04747},
 primaryClass = {astro-ph.GA},
       adsurl = {https://ui.adsabs.harvard.edu/abs/2017ApJ...842...33V}
}

@ARTICLE{Walmsley1980,
       author = {{Walmsley}, C.~M. and {Winnewisser}, G. and {Toelle}, F.},
        title = "{Cyanoacetylene and cyanodiacetylene in interstellar clouds}",
      journal = {\aap},
         year = 1980,
        month = jan,
       volume = {81},
       number = {1-2},
        pages = {245-250},
       adsurl = {https://ui.adsabs.harvard.edu/abs/1980A&A....81..245W}
}

@ARTICLE{Walmsley2004,
       author = {{Walmsley}, C.~M. and {Flower}, D.~R. and {Pineau des For{\^e}ts}, G.},
        title = "{Complete depletion in prestellar cores}",
      journal = {\aap},
         year = 2004,
        month = may,
       volume = {418},
        pages = {1035-1043},
          doi = {10.1051/0004-6361:20035718},
archivePrefix = {arXiv},
       eprint = {astro-ph/0402493},
 primaryClass = {astro-ph},
       adsurl = {https://ui.adsabs.harvard.edu/abs/2004A&A...418.1035W}
}

@ARTICLE{WardThompson1999,
       author = {{Ward-Thompson}, D. and {Motte}, F. and {Andre}, P.},
        title = "{The initial conditions of isolated star formation - III. Millimetre continuum mapping of pre-stellar cores}",
      journal = {\mnras},
         year = 1999,
        month = may,
       volume = {305},
       number = {1},
        pages = {143-150},
          doi = {10.1046/j.1365-8711.1999.02412.x},
       adsurl = {https://ui.adsabs.harvard.edu/abs/1999MNRAS.305..143W}
}

@ARTICLE{HC3Nref,
       author = {{Mallinson}, P.~D. and {de Zafra}, Robert L.},
        title = "{The microwave spectrum of cyanoacetylene in ground and excited vibrational states}",
      journal = {Molecular Physics},
         year = 1978,
        month = jan,
       volume = {36},
       number = {3},
        pages = {827-843},
          doi = {10.1080/00268977800101971},
       adsurl = {https://ui.adsabs.harvard.edu/abs/1978MolPh..36..827M}
}

@ARTICLE{H13C3Nref,
       author = {{Creswell}, R.~A. and {Winnewisser}, G. and {Gerry}, M.~C.~L.},
        title = "{Rotational spectra of the $^{13}$C and $^{15}$N isotopic species of cyanoacetylene.}",
      journal = {Journal of Molecular Spectroscopy},
         year = 1977,
        month = jan,
       volume = {65},
        pages = {420-429},
          doi = {10.1016/0022-2852(77)90281-8},
       adsurl = {https://ui.adsabs.harvard.edu/abs/1977JMoSp..65..420C}
}

@ARTICLE{CH3CCHref,
journal = {C. R. Acad. Sci. Paris},
title = {},
volume = {B 268},
pages = {800},
year = {1969},
author = {{Bauer}, A and {Burie}, J},
}

@article{CH3CCD_CH2DCCHref,
title = {Rotational Spectra of CH2DCCH and CH3CCD: Experimental and ab Initio Equilibrium Structures of Propyne},
journal = {Journal of Molecular Spectroscopy},
volume = {160},
number = {2},
pages = {471-490},
year = {1993},
issn = {0022-2852},
doi = {https://doi.org/10.1006/jmsp.1993.1194},
url = {https://www.sciencedirect.com/science/article/pii/S002228528371194X},
author = {M. Leguennec and J. Demaison and G. Wlodarczak and C.J. Marsden},
}

@ARTICLE{CNroute,
       author = {{Sims}, Ian R. and {Queffelec}, Jean-Louis and {Travers}, Daniel and {Rowe}, Bertrand R. and {Herbert}, Lee B. and {Karth{\"a}user}, Joachim and {Smith}, Ian W.~M.},
        title = "{Rate constants for the reactions of CN with hydrocarbons at low and ultra-low temperatures}",
      journal = {Chemical Physics Letters},
         year = 1993,
        month = aug,
       volume = {211},
       number = {4-5},
        pages = {461-468},
          doi = {10.1016/0009-2614(93)87091-G},
       adsurl = {https://ui.adsabs.harvard.edu/abs/1993CPL...211..461S}
}

@ARTICLE{HCNroute,
       author = {{Iraqi}, M. and {Petrank}, A. and {Peres}, M. and {Lifshitz}, C.},
        title = "{Proton transfer reactions of C2H+2 : the bond energy D0 (C2HH)}",
      journal = {International Journal of Mass Spectrometry and Ion Processes},
         year = 1990,
        month = oct,
       volume = {100},
        pages = {679-691},
          doi = {10.1016/0168-1176(90)85102-8},
       adsurl = {https://ui.adsabs.harvard.edu/abs/1990IJMSI.100..679I}
}

@ARTICLE{HNCroute,
       author = {{Fukuzawa}, Kaori and {Osamura}, Yoshihiro},
        title = "{Molecular Orbital Study of Neutral-Neutral Reactions concerning HC$_{3}$N Formation in Interstellar Space}",
      journal = {\apj},
         year = 1997,
        month = nov,
       volume = {489},
       number = {1},
        pages = {113-121},
          doi = {10.1086/304782},
       adsurl = {https://ui.adsabs.harvard.edu/abs/1997ApJ...489..113F}
}
%

\FloatBarrier

\begin{appendix}
\onecolumn

\section{Additional material on deuteration maps}

\subsection{Uncertainties of deuteration maps}

\begin{figure}[h]
    \includegraphics[width=0.49\textwidth]{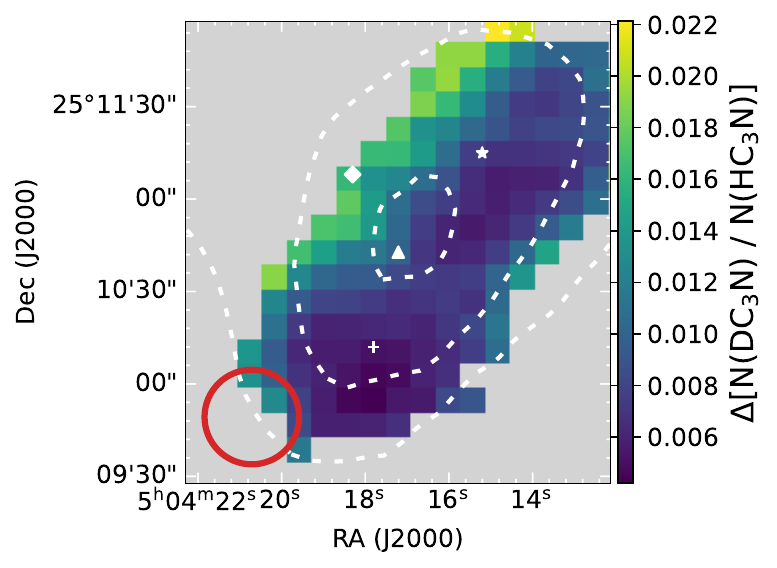}
    \includegraphics[width=0.49\textwidth]{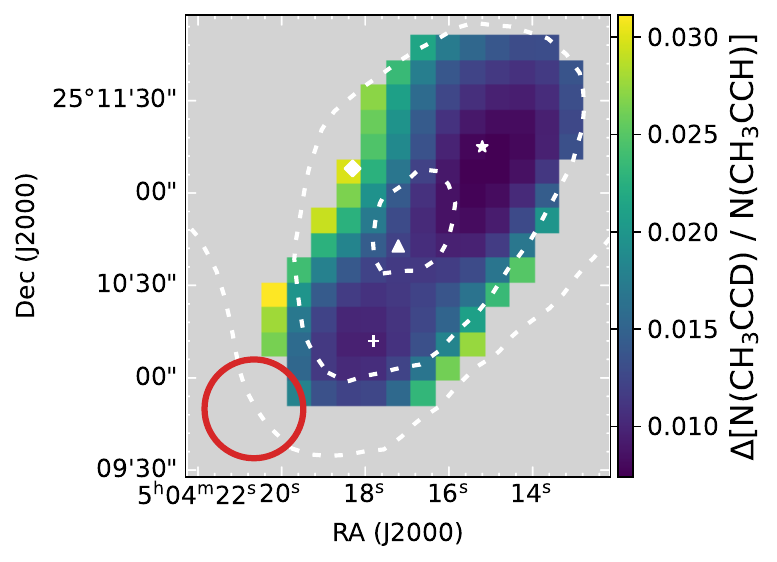}
    \includegraphics[width=0.49\textwidth]{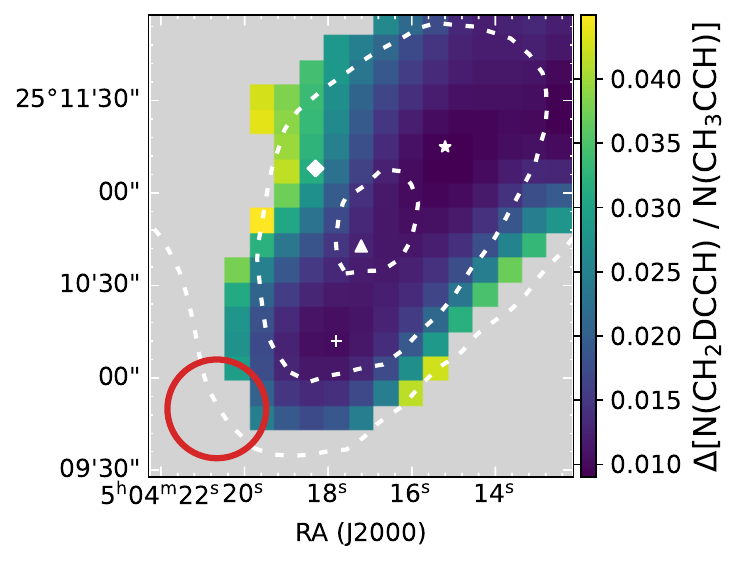}
    \includegraphics[width=0.49\textwidth]{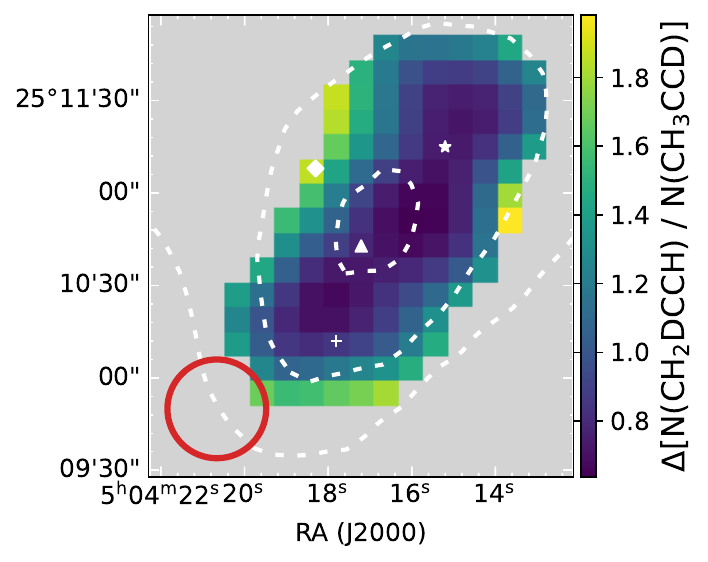}
    \caption{Error maps of the deuteration maps for \cyano{} and \prop{} (presented in Figs.~\ref{fig:deuterationmapsHC3N} and \ref{fig:deuterationmapsCH3CCH}, respectively). Only pixels above the $3\sigma$ level of the respective integrated intensities are plotted.
    The dashed line contours represent 30\%, 50\%, and 90\% of the H$_2$ column density peak derived from $Herschel$ maps \citep{Spezzano2016}. The circle in the bottom-left corner indicates the beam size of the IRAM 30\,m telescope (31"). The markers in white represent the dust peak (triangle) and the molecular emission peaks of \meth{} (diamond), \prop{} (star), and \cyc{} (plus sign).}
    \label{fig:DeuterationErrormaps}
\end{figure}

Figure~\ref{fig:DeuterationErrormaps} shows the uncertainties on the deuteration maps of \cyano{} and \prop{} presented in Figs.~\ref{fig:deuterationmapsHC3N} and \ref{fig:deuterationmapsCH3CCH}, respectively.

\subsection{Deuteration of \prop{} derived from K=1}\label{sec:app:DeuterationOfCH3CCH}

\begin{figure}[h]
    \includegraphics[width=0.33\textwidth]{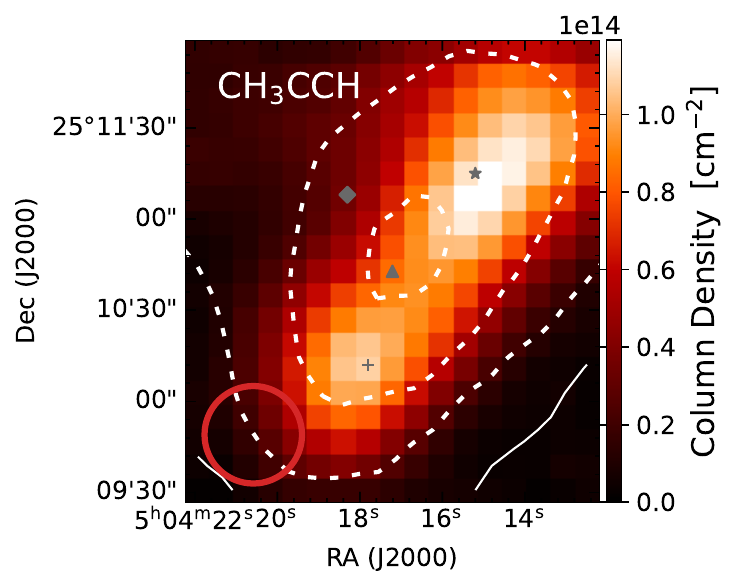}
    \includegraphics[width=0.33\textwidth]{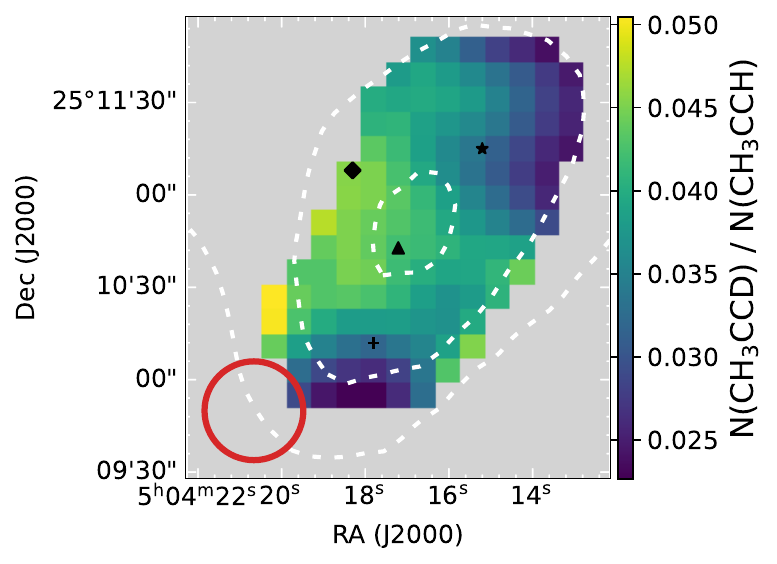}
    \includegraphics[width=0.33\textwidth]{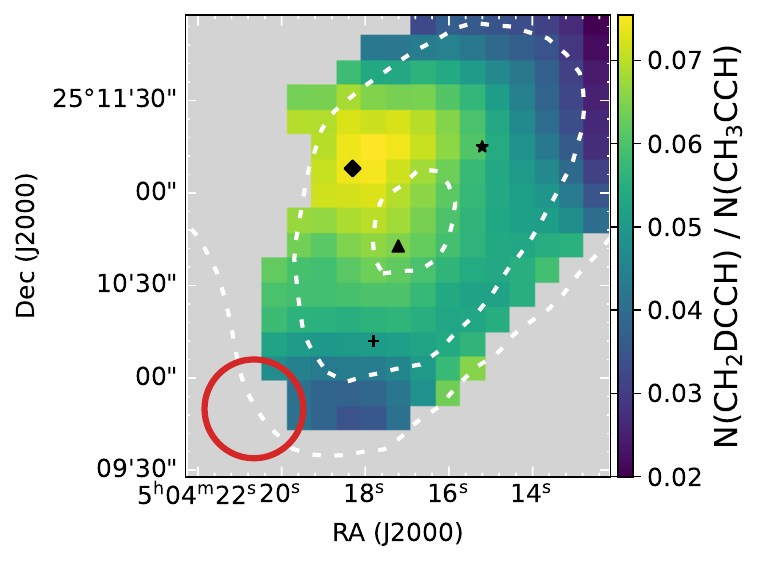}
    \caption{\textit{Left:} Column density map of \prop{} derived with the ($5_1-4_1$) transition. The solid contour indicates the 3$\sigma$ level of the column density. The markers represent the dust peak (triangle) and the molecular emission peaks of \meth{} (diamond), \prop{} (star), and \cyc{} (plus sign). \textit{Right:} Corresponding deuteration maps. Only pixels above the $3\sigma$ level of the respective integrated intensities are plotted.}
    \label{fig:DeuterationOfCH3CCH5141}
\end{figure}

Figure~\ref{fig:DeuterationOfCH3CCH5141} shows the column density map of \prop{} derived from the ($5_1-4_1$) transition and the resulting deuteration maps. 
The distribution of the emission of the K=1 transition is identical to the K=0 transition, resulting in an identical morphology the deuterium fraction.
However, the peak column density derived from K=1 with an excitation temperature of 10\,K is a factor of 2 higher compared to the values derived from K=0. 
This results in substantially lower deuteration levels, with peak values of 0.045(5) and 0.07(1) for N(\propd{})/N(\prop{}) and N(\dprop{})/N(\prop{}), respectively.
However, the relation between the two deuterated isotopologues does not change, and therefore, the conclusions drawn from them are not affected.

\section{Spectra at molecular peaks}

\begin{figure*}[h]
    \includegraphics[width=\textwidth]{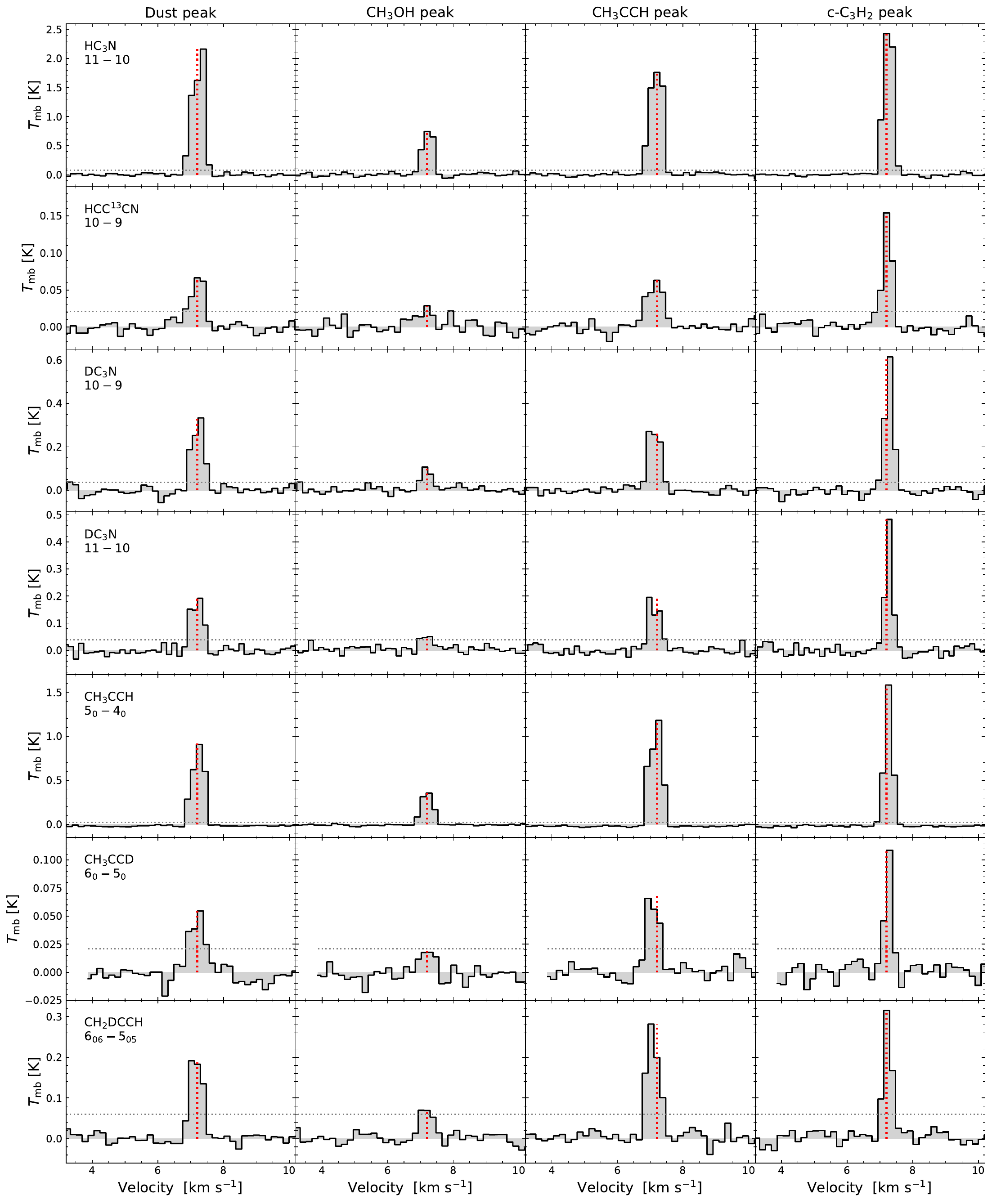}
    \caption{Spectra of the observed transitions, extracted at the dust peak, the \meth{} peak, the \prop{} peak, and the \cyc{} peak of L1544. The extraction locations of the spectra, as well as the telescope beam used as aperture, are indicated in Fig.~\ref{fig:IntegratedIntensityMaps}. The horizontal line indicates the 3$\sigma$ confidence level, and the vertical line indicates the systemic velocity of the source.}
    \label{fig:peakspectra}
\end{figure*}

Figure~\ref{fig:peakspectra} presents spectra of the observed transitions, extracted towards the three molecular emission peaks (\meth{}, \prop{}, \cyc{}) and the dust peak of L1544, by using a circular aperture with a diameter corresponding to the telescope beam size (31"). 
\prop{} and \dprop{} are only represented with one transition each, as they are discussed separately in Sect.~\ref{sec:analysis} and Fig.~\ref{fig:CH3CCHmoleculepeaks}.
The extraction locations of the spectra are marked in Fig.~\ref{fig:IntegratedIntensityMaps}.
The corresponding linewidths and velocities of the spectra at the four extraction locations are derived with Gaussian fitting and given in Table~\ref{Tab:MolecularPeakLinewidths}.

The spectra show that the brightest emission of all observed molecules is detected towards the \cyc{} peak located in the south of the core. In this region, the core is less protected from the ISRF, which drives photochemistry and leads to an active carbon chemistry and an increased abundance of C-bearing species. As all observed molecules in this study are essentially carbon chains, they are most efficiently formed in such conditions.
However, the linewidths observed towards the \prop{} peak (and the dust peak) are significantly larger than towards the \cyc{} peak. This indicates the location where the flow of material from the outer regions of the core (in the direction of the main filament) reaches the dense core and undergoes a local slow shock (as in the case of \meth{} at the \meth{} peak, see \citealt{Lin2022}), causing broader linewidths.
The spectra of \dcyano{} seem to show rather flat line profiles towards the \prop{} peak. As \dcyano{} is tracing less dense gas in the outer envelope of the core, the broader linewidth is likely caused by the larger turbulence present in those outer parts \citep[e.g.][]{FullerMyers1992}.

\begin{table}[h]
    \centering
    \caption{Comparison of the linewidths and velocities of the observed transitions at different locations in L1544 (dust peak, \meth{} peak, \prop{} peak, and \cyc{} peak, see Fig.~\ref{fig:IntegratedIntensityMaps}). }
    \begin{tabular}{l c l l l l l l l l}
    \hline\hline 
    \noalign{\smallskip}
    Molecule & Transition & \multicolumn{2}{c}{Dust peak} & \multicolumn{2}{c}{\meth{} peak} & \multicolumn{2}{c}{\prop{} peak} & \multicolumn{2}{c}{\cyc{} peak} \\
     &  & $V_\mathrm{LSR}$ & FWHM  & $V_\mathrm{LSR}$ & FWHM & $V_\mathrm{LSR}$ & FWHM  & $V_\mathrm{LSR}$ & FWHM \\ 
    \noalign{\smallskip}
    \hline
    \noalign{\smallskip}
    \cyano{} & $11-10$ & 7.250(7) & 0.197(7) & 7.239(5) & 0.180(5) & 7.200(6) & 0.204(6) & 7.268(3) & 0.152(3) \\
    \xcyano{} & $10-9$ & 7.19(2)  & 0.21(2)  & 7.17(4)  & 0.19(4)  & 7.13(2)  & 0.24(2)  & 7.239(6) & 0.141(6) \\
    \dcyano{} & $10-9$ & 7.22(1)  & 0.19(1)  & 7.18(2)  & 0.18(2)  & 7.141(9) & 0.203(9) & 7.285(4) & 0.130(3) \\
    \prop{}  & $5_0-4_0$ & 7.24(3) & 0.19(3) & 7.19(2)  & 0.18(2)  & 7.17(3)  & 0.20(3)  & 7.26(2)  & 0.13(2) \\
    \propd{} & $6_0-5_0$ & 7.17(4) & 0.25(4) & 7.18(5)  & 0.21(5)  & 7.07(3)  & 0.20(3)  & 7.26(2)  & 0.12(1) \\
    \dprop{} & $6_{06}-5_{05}$ & 7.151(9) & 0.195(9) & 7.10(2) & 0.18(2) & 7.062(8)  & 0.193(8)  & 7.229(6) & 0.136(5) \\
    \noalign{\smallskip}
    \hline
    \end{tabular}  
    \tablefoot{The parameters are derived from the spectra presented in Fig.~\ref{fig:peakspectra} using Gaussian fitting, and given in units of km\,s$^{-1}$.}
    \label{Tab:MolecularPeakLinewidths}
\end{table}

\end{appendix}

\end{document}